\documentclass[preprint,nofootinbib,aps,11pt]{revtex4-1}
\usepackage{graphicx}
\usepackage{dcolumn}
\usepackage{bm}
\usepackage{epsfig,amsmath}
\usepackage{amssymb}
\usepackage{subfigure,esint}
\usepackage{float}
\usepackage[usenames,dvipsnames]{color}
\usepackage[pagebackref=false, colorlinks=true]{hyperref}
\definecolor{redish}{rgb}{0.7,0.2,0.0}  
\definecolor{bluish}{rgb}{0.2,0.5,0.8}

\hypersetup{linkcolor=redish,          
                  citecolor=blue,        
                  filecolor=magenta,      
                  urlcolor=blue}          

\DeclareFontFamily{U}{rsfs}{}         
\DeclareFontShape{U}{rsfs}{m}{n}{<5> rsfs5 <6><7> rsfs7          %
  <8><9><10><10.95><12><14.4><17.28><20.74><24.88> rsfs10}{}     %
\DeclareMathAlphabet{\mathfs}{U}{rsfs}{m}{n}

\begin{document}
\author{Suvankar Paul}
\email{svnkr@iitk.ac.in}
\author{Rajibul Shaikh}
\email{rshaikh@iitk.ac.in}
\author{Pritam Banerjee}
\email{bpritam@iitk.ac.in}
\author{Tapobrata Sarkar}
\email{tapo@iitk.ac.in}
\affiliation{Department of Physics, \\ Indian Institute of Technology, \\ Kanpur 208016, India}

\title{\Large Observational signatures of wormholes with thin accretion disks}

\bigskip

\begin{abstract}
We numerically construct images of thin accretion disks in rotating wormhole backgrounds, for the 
Kerr-like and the Teo class of wormholes. Our construction is illustrated by two methods, a semi-analytic 
scheme where separated null geodesic equations obtained by analytically integrating the second order equations once are used, 
as well as by a numerical ray-tracing method utilizing a fourth order Runge-Kutta 
algorithm. Our result shows dramatic differences between accretion disk images in wormhole 
backgrounds, compared to black hole ones, specifically because a wormhole can in
principle have accretion disks on both sides of its throat. We establish the nature of the images
if the observer and the disk are on two opposite sides of the throat, and show that these can provide 
conclusive observational evidence of wormhole geometries.

\end{abstract}
\maketitle


\section{Introduction}

It is believed that supermassive black holes with masses of the order of $10^6~-~10^{10}M_{\odot}$ exist at the 
center of most galaxies. Observational aspects of black holes, which are characterized by an event 
horizon, are naturally of great significance, as these might shed light on the underlying physics of the end stages of
gravitational collapse, which are perhaps influenced by quantum phenomena in the regimes of very strong gravity. 
Indeed, recent observations of the center of the galaxy M87 by the Event Horizon Telescope \citep{EHT1,EHT2,EHT3}
have triggered a flurry of activities towards understanding and quantifying possible images of accretion disks 
that can be formed around the galactic center. 

While the notion of the event horizon continues to attract much attention, it has become clear by now that 
compact objects which do not have an event horizon might mimic many of the properties of black holes
(for a review of the topic, we refer the reader to the recent work of Cardoso and Pani \citep{CardosoReview}). 
One such object is the wormhole - an exotic solution of Einstein's equations that connect two different
universes or two distant regions of the same universe by a throat. Celestial objects that are not tidally disrupted can tunnel through the throat,
from one universe or one distant region to another. 
An important issue related to this tunneling of material objects through the wormhole throat
is that of traversability, and it was shown by Fuller and Wheeler \citep{FullerWheeler} 
that the Schwarzchild wormhole (also called the Einstein-Rosen bridge) is not traversable in this sense. 
Traversable wormholes, which are of physical interest appeared first in the work of Morris and Thorne \citep{MorrisThorne} 
and a subsequent description of a ``time machine'' based on the Moris-Thorne construction 
appeared in \citep{MTY,Novikov}. Details of these can be found in the excellent book by Visser \citep{VisserBook} 
(see also \citep{Lemos} for more recent related literature). Typically, the matter sourcing wormhole geometries violates standard energy conditions \citep{MorrisThorne}. However, several attempts have been made 
to evade such violations, and as is well known, dynamical scenarios \citep{Sayan1} or wormhole geometries in
modified gravity \citep{WMG1,WMG2,WMG3,WMG4,WMG5,WMG6,WMG7,WMG8,WMG9, WMG10} 
may offer situations in which these energy conditions are not violated. Stability of
wormhole geometries which is known to be related to the equation of state of the matter supporting such geometries, 
is also a much studied topic, although the issue is still debated. While it is known that
some wormhole geometries might be stable under metric and field perturbations \citep{stab0, stab0a, stab0b},
others might not \citep{stab1,stab2,stab3}. However, in spite of these issues, wormholes
continue to attract much attention as they are prototype examples of solutions of general relativity that can
mimic black holes.

The fact that wormholes can strongly resemble black holes 
was noted more than a decade ago by Damour and Solodukhin \citep{DamourSolodukhin}, who pointed out that
several features of black holes such as quasi-normal modes, accretion properties, no-hair theorems etc. can
be closely mimicked by wormhole geometries as well. Indeed, after the first results on gravitational wave 
detection by LIGO \citep{LIGO} were obtained, it was shown by Cardoso, Franzin and Pani \citep{Cardoso1} that a class of
wormholes that have a thin shell of phantom matter at the throat can exhibit an initial quasinormal ringdown mode that is entirely
similar to that of a black hole, with differences emerging only at late times. Later, Konoplya and Zhidenko \citep{KZ} showed 
that specific classes of wormholes can in fact ring similarly or differently compared to black holes at all times. 
In the light of the above discussions, it is clearly important and interesting to further study the observational 
distinctions between black hole and wormhole geometries. For a sampling of recent literature on the topic, we
refer the reader to \cite{bambi_2013a,nedkova_2013,ohgami_2015, mustafa_2015,abdujabbarov_2016b,rajibul_2018b,gyulchev_2018, amir_2018,jusufi_2018,shaikh_2017,rajibul_2018c,shaikh_2019a,Dai}\footnote{See \cite{Mizuno} for recent results
on testing alternative theories of gravity via an analysis of black hole shadows.}.

In this paper, we point out that one can observe dramatic differences between these objects in the context of their
accretion disk images (see \cite{disk1,disk2,disk3,disk4,disk5,disk6} for some works on images of accretion disks around different objects). We proceed with two assumptions here. First, we will restrict ourselves to thin accretion
disks in wormhole geometries, which were analytically studied in \cite{harko_2009}. Second, we will assume
that the accreting matter does not interact with the matter seeding the wormhole, which is a fairly standard 
assumption in the literature. 
Specifically, since the wormhole has a throat, it is possible and indeed natural to have accretion disks on either or both 
sides of the wormhole throat. If one only focuses on the disk on the same side of the throat as that of the observer, then
the images may or may not mimic those of black hole accretion disks. However, we show here 
that the images of the disk when it is on the side of the throat opposite to that of the observer are drastically different from the ones observed from black holes. In a related context, we have shown 
recently \citep{shaikh_2019a} that strong gravitational lensing from wormhole geometries might be qualitatively 
different from the ones in black hole backgrounds. The reason for this is two-fold. Namely, with the observer and
the source both on the same side of the wormhole throat, the throat can itself act as an effective photon sphere, where light 
travels in unstable circular orbits and via a small perturbation can reach an observer at infinity. Also, the observer
and the source might be on different sides of the wormhole throat which opens up a further possible feature of 
gravitational lensing that is absent in black hole geometries. Here, we logically continue this analysis further, and 
focus on the images of accretion disks of black hole and wormhole geometries. 

\begin{figure}[ht]
\centering
\subfigure[~Kerr black hole]{\includegraphics[scale=0.60]{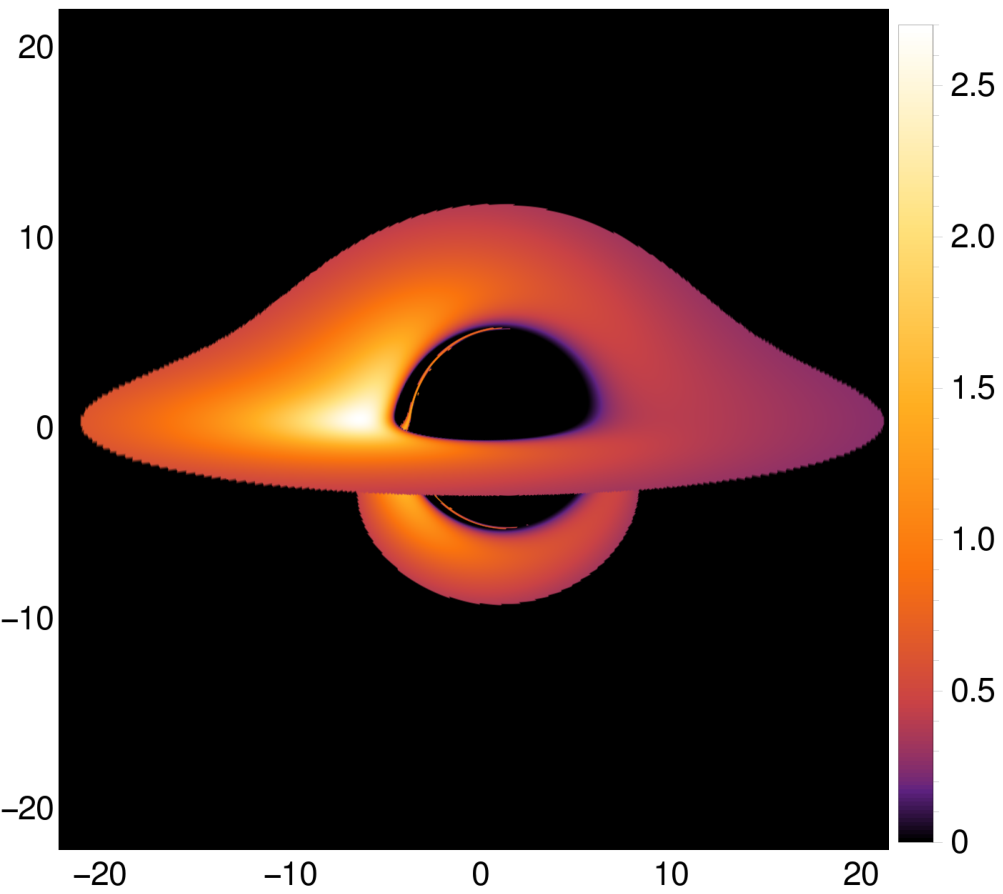}\label{fig:IntroA}}\hspace{0.1cm}
\subfigure[~Kerr like wormhole]{\includegraphics[scale=0.60]{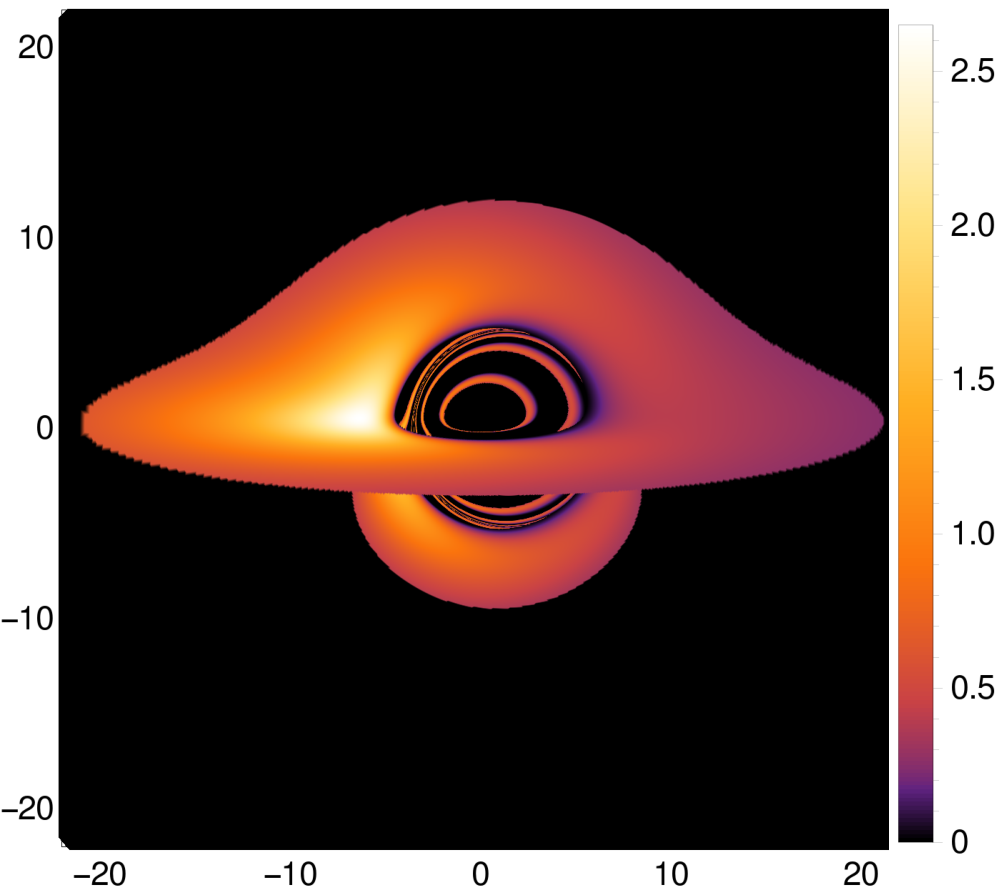}\label{fig:IntroB}}\hspace{0.3cm}
\caption{Typical intensity maps for accretion disks from (a) the Kerr black hole and (b) the Kerr-like wormhole that has accretion
disks on both sides of the throat.}
\label{figintro}
\end{figure}

In this work, we will consider two classes of rotating wormhole geometries. The first is the Kerr-like wormhole constructed 
by Bueno et. al. in \cite{bueno_2018} and the second is the rotating version of the Morris-Thorne wormhole, also known as the
Teo wormhole \citep{teo_1998}. We construct accretion disk images for these wormholes numerically. 
Two different numerical algorithms are considered here. The first one is a semi-analytic construction, in which 
separated null geodesic equations obtained by analytically integrating the second order equations once are used, 
and the second,
in which a fully numerical ray-tracing method is used by employing a fourth order Runge-Kutta algorithm. 
With both these methods, we obtain the accretion disk images of the two classes of wormhole geometries
considered here. Importantly, considering the fact that there might be accretion disks on both sides
of the wormhole throat, we get an overall picture that is strikingly different from any black hole accretion disk image
obtained so far in the literature. This is the main contribution of this work. To set the stage, we reproduce typical images
obtained from our analysis, of the Kerr black hole and the Kerr-like wormhole with accretion disks on both sides of the throat, 
in Fig. (\ref{figintro}), with parameter values that will be discussed in section (\ref{sec:images}). 
The marked qualitative differences in the intensity maps are immediately apparent. We expect these 
features of the wormhole images to be generic, and the formalism for obtaining these will be discussed in sequel. 

This paper is organized as follows. In the next section \ref{sec:intro}, we first record the expressions for the wormhole metrics, 
and then proceed to recapitulate some basic facts about accretion disks around wormholes and black holes. 
Section \ref{sec:tracing} deals with the numerical methods to construct the images of these accretion disks. The two different
numerical methods mentioned in the previous paragraph are elaborated in detail here. Section \ref{sec:images} contains
the main results of this paper, where we show the intensity maps of the accretion disk images for the two classes of
wormholes considered here, and we comment on the novel features that we obtain by our analysis. 
Section \ref{sec:conclusions} ends this paper with our broad conclusions.

\section{Basic formalism}\label{sec:intro}

In this section, we review the basic ingredients needed for the construction of accretion disk images in wormhole
geometries. We first specify the space-time metrics, and then record some known analytical results regarding
accretion disks in these backgrounds. This section has review material, meant to set up the scenarios to be used later
in this paper. 

\subsection{Wormhole spacetimes}
\label{sec:spacetimes}

In this work, we consider two different rotating wormhole spacetimes, namely the Kerr-like and the Teo wormhole. 
The stationary and axi-symmetric spacetime geometry of the Kerr-like wormhole was obtained in \citep{bueno_2018},
and is given by
\begin{eqnarray}
ds^2&=&-\left(1-\frac{2M r}{\Sigma}\right)dt^2-\frac{4Mar\sin^2\theta}{\Sigma
}dtd\phi+\frac{\Sigma}{\hat \Delta}dr^2 +\Sigma d\theta^2+\left(r^2+a^2+
\frac{2Ma^2r\sin^2\theta}{\Sigma}\right)\sin^2\theta d\phi^2~,\nonumber\\
\Sigma &\equiv& r^2+a^2\cos^{2}{\theta},\quad \hat\Delta \equiv r^{2} - 2M(1+\lambda^2)r + a^{2}.
\label{eq:KerrWH}
\end{eqnarray}
The parameters $a$ and $M$ are related, respectively, to the spin and mass of the wormhole, and $\lambda$ 
is the deformation parameter whose value signifies the deviation of the metric of Eq. (\ref{eq:KerrWH}) 
from the Kerr black hole. For $\lambda=0$, we recover the 
Kerr black hole and the corresponding black hole horizons are given by $\hat{\Delta}(r_{\pm})=0$, i.e., by 
$r_{\pm}=M\pm \sqrt{M^2-a^2}$. For $\lambda\neq 0$, however, the spacetime represents a wormhole and the wormhole throat $r_{th}$ is given 
by $\hat{\Delta}(r_{th})=0$, .i.e., by
\begin{equation}
r_{th}=M(1+\lambda^2)+\sqrt{M^2(1+\lambda^2)^2-a^2}.
\label{modifiedth}
\end{equation}
Note that Eq. (\ref{modifiedth}) necessarily implies that $r_{th}>r_{+}$, $r_{+}$ being the outer horizon of the Kerr black hole.

Our second example would be the stationary, axisymmetric spacetime metric describing a rotating traversable wormhole 
of the Teo class, and is given by the metric \citep{teo_1998} 
\begin{equation}
ds^2=-N^2 dt^2+\frac{dr^2}{1-\frac{b}{r}}+r^2K^2
\left[d\theta^2+\sin^2\theta(d\varphi-\omega dt)^2
\right],
\label{eq:teo_wormhole}
\end{equation}
where $-\infty<t<\infty$, and $r_{th}\leq r<\infty$, $0\leq\theta\leq\pi$ and $0\leq\phi\leq 2\pi$ are spherical 
coordinates. The functions $N$, $b$, $K$, and $\omega$ depend on $r$ and $\theta$ only, such that it is regular 
on the symmetry axis $\theta=0,\pi$ \citep{teo_1998}. The spacetime describes two identical, asymptotically flat 
regions connected together at the throat $r=r_{th}=b>0$. The above metric is a rotating generalization of the static 
Morris-Thorne wormhole \citep{MorrisThorne}. In this work, we consider the following forms of the metric functions 
for the Teo wormhole, which are frequently used in the literature,
\begin{equation}
N^2 =\exp\left[-\frac{r_{th}}{r}\right], \quad b(r)=r_{th}=2M~,~~
K =1, \quad \omega = \frac{2J}{r^3},
\label{eq:metric_choice}
\end{equation}
where $J$ and $M$ are respectively the angular momentum and mass of the wormhole, and $r_{th}=2M$ 
is the wormhole throat. The spin of the wormhole is given by $a=J/M$.

The above wormholes which we consider in this work are symmetric wormholes, i.e., wormholes whose two 
sides of the throat are represented by the same copy of the spacetime. However, for asymmetric wormholes, 
the spacetimes on the two sides are different. In this work, we do not consider this latter case for simplicity. 
Sometimes, a wormhole spacetime is written in the proper radial coordinate $l$ (see \cite{rajibul_2018c}) or a coordinate $x$ (see \cite{bronnikov_2019}) such 
that $r=r(l)$ or $r=r(x)$ has a minimum at $l=0$ or $x=0$ and the minimum value corresponds to the throat 
radius $r_{th}$. In such a case, $l>0$ or $x>0$ represents one side and $l<0$ or $x<0$ represents 
another side of the throat. As we shall discuss, writing the metric in this coordinate ($l$ or $x$) may sometimes 
become useful when we consider scenarios involving both sides. For ease of discussion in the subsequent 
sections, we write the stationary, axially symmetric rotating wormhole spacetimes in the form
\begin{equation}
ds^2=g_{tt}dt^2+2g_{t\phi}dtd\phi+g_{rr}dr^2+g_{\theta\theta}d\theta^2+ g_{\phi\phi} d\phi^2,
\label{eq:general_metric}
\end{equation}
where the metric functions $g_{\mu\nu}$, in general, depend on $r$ and $\theta$. In the equatorial ($\theta=\pi/2$) plane, 
however, they depend on the radial coordinate $r$ only.

\subsection{Accretion disks around wormholes and black holes}
\label{sec:disks}
A geometrically thin accretion disk consists of massive particles moving in stable circular timelike geodesics. 
We consider the Novikov-Thorne model of a thin accretion disk \citep{novikov_1973,page_1974}. Since we are 
dealing with stationary and axially symmetric spacetimes of the form given in Eq. (\ref{eq:general_metric}), there are two constants 
of motion along a timelike geodesic, namely, the specific energy $\tilde{E}$ (energy per unit mass) and the specific 
angular momentum $\tilde{L}$ of a massive particle. Therefore, the timelike geodesic equations on the equatorial plane 
are given by 
\begin{equation}
\dot{t}=\frac{\tilde{E} g_{\phi\phi}+\tilde{L}g_{t\phi}}{g_{t\phi}^2-g_{tt}g_{\phi\phi}}, \quad \dot{\phi}
=-\frac{\tilde{E} g_{t\phi}+\tilde{L}g_{tt}}{g_{t\phi}^2-g_{tt}g_{\phi\phi}}~,~~
g_{rr}\dot{r}^2=-1+\frac{\tilde{E}^2 g_{\phi\phi}+2\tilde{E}\tilde{L}g_{t\phi}+\tilde{L}^2g_{tt}}{g_{t\phi}^2-g_{tt}g_{\phi\phi}}=\tilde{V}_{eff}(r),
\label{eq:timelike_geodesics}
\end{equation}
where an overdot represents differentiation with respect to the affine parameter (unless otherwise specified), 
and $\tilde{V}_{eff}$ is the 
effective potential. A stable circular timelike geodesic satisfies $\dot{r}=0$, $\ddot{r}=0$ and $\dddot{r}<0$. 
For a stable circular timelike geodesic which lies outside the wormhole throat, i.e., the stable circular orbit 
radius $r>r_{th}$, in terms of the effective potential, the conditions become $\tilde{V}_{eff}=0$, $\tilde{V}'_{eff}=0$ 
and $\tilde{V}''_{eff}<0$. The first two conditions yield the specific energy and the specific angular momentum of 
the particles moving in the stable circular orbits. These are given by \citep{harko_2009}
\begin{equation}
\tilde{E}=-\frac{g_{tt}+g_{t\phi}\Omega}{\sqrt{-g_{tt}
-2g_{t\phi}\Omega-g_{\phi\phi}\Omega^2}}~,~~
\tilde{L}=\frac{g_{t\phi}+g_{\phi\phi}\Omega}{\sqrt{-g_{tt}
-2g_{t\phi}\Omega-g_{\phi\phi}\Omega^2}}~,~~
\Omega=\frac{d\phi}{dt}=\frac{-g_{t\phi}'+\sqrt{(g_{t\phi}')^2
-g_{tt}'g_{\phi\phi}'}}{g_{\phi\phi}'},
\label{eq:omega}
\end{equation}
where a prime denotes a derivative with respect to $r$, and $\Omega=d\phi/dt$ is the angular momentum of 
the particles forming the disk. The flux of the electromagnetic radiation emitted from a radial position $r$ of a 
disk is given by the standard formula \citep{novikov_1973,page_1974}
\begin{equation}
\mathcal{F}(r)=-\frac{\dot{M}}{4\pi\sqrt{-g}}\frac{\Omega'}{(\tilde{E}-\Omega \tilde{L})^2}\int_{r_{in}}^r (\tilde{E}-\Omega \tilde{L})\tilde{L}' dr,
\label{eq:flux}
\end{equation}
where $\dot{M}=dM/dt$ is the mass accretion rate, and $r_{in}$ is the inner edge of the disk. Also,
$\sqrt{-g}=\sqrt{g_{rr}(g_{t\phi}^2-g_{tt}g_{\phi\phi})}$ is the determinant of the metric on the equatorial plane.

The inner edge of an accretion disk is generally located at the marginal stable orbit $r=r_{ms}$, given by $\dddot{r}=0$. 
For a wormhole, if the marginal stable orbit is outside the throat, i.e., if $r_{ms}>r_{th}$, then, in terms of the effective 
potential, the marginal stable orbit condition becomes $\tilde{V}_{eff}''(r_{ms})=0$. However, if a real $r_{ms}>r_{th}$ does 
not exist, then there is no marginal orbit outside the throat. In such a case, if circular orbits, which satisfy 
$\tilde{V}_{eff}(r)=0$, $\tilde{V}'_{eff}(r)=0$ and $\tilde{V}''_{eff}(r)<0$, exist upto the throat, then the circular orbit 
at the throat will act as the marginal stable orbit. The reason for this is as follows. Note that $g_{rr}^{-1}=0$ at the 
throat $r=r_{th}$, and hence, $\dot{r}=0$ is automatically satisfied at $r=r_{th}$. Therefore, $\tilde{V}_{eff}=0$ 
and $\tilde{V}'_{eff}=0$ at $r=r_{th}$, respectively, implies that $\ddot{r}=0$ and $\dddot{r}=0$. This is in contrast to the 
stable circular orbits outside the throat for which $\tilde{V}_{eff}=0$ and $\tilde{V}'_{eff}=0$ means 
$\dot{r}=0$ and $\ddot{r}=0$, as $g_{rr}^{-1}$ does not vanish outside the throat. Therefore, if there is no marginal 
stable orbit outside the throat and stable circular orbits exist all the way upto the throat, then the stable circular 
orbit at the throat acts as the marginal stable orbit.

Having discussed the basic ingredients needed for our analysis, we will now proceed to specify 
some details of our numerical computation scheme. 
 
\section{Tracing the observed rays}
\label{sec:tracing}
The photons emitted from an accretion disk undergo gravitational lensing and a fraction of them reach a distant observer. 
Therefore, in order to obtain the intensity map of the image produced in the observer's sky, we need to trace back the 
observed photons. For this, we integrate the null geodesic equations backward in time from the 
observer's position $(r_o,\theta_o)$ to the emission point $(r_e,\pi/2)$ on the accretion disk. We do 
this in two ways by a semi-analytic and a full numerical method. These are discussed in the following subsections.

\subsection{Semi-analytic ray-tracing}
This technique can be applied in spherically or stationary axially symmetric spacetimes which allow separation of the 
null geodesic equations using the constants of motion such as energy $E$, angular momentum $L$ of the photons and 
the Carter constant $\mathcal{K}$. This means that the second-order geodesic equations can be integrated once to obtain 
a set of first order geodesic equations giving $\dot{x}^\mu=\{\dot{t},\dot{r},\dot{\theta},\dot{\phi}\}$, the constants of 
motion $E$, $L$ and $\mathcal{K}$ being the integration constants. This can be done in a large class of spacetimes. For example, see \cite{rajibul_2019} where it is shown that the null geodesic equations in an arbitrary stationary and axially symmetric spacetime obtained through the Newman-Janis algorithm can be completely separated to obtain a set of first order geodesic equations. In any stationary and axially symmetric spacetime of the form given in Eq. (\ref{eq:general_metric}), the geodesic equations for $\dot{t}$ and $\dot{\phi}$ can be written as
\begin{equation}
\dot{t}=\frac{E g_{\phi\phi}+L g_{t\phi}}{g_{t\phi}^2-g_{tt}g_{\phi\phi}}, \quad \dot{\phi}
=-\frac{E g_{t\phi}+L g_{tt}}{g_{t\phi}^2-g_{tt}g_{\phi\phi}}.
\label{eq:t-phi}
\end{equation}
Note that the above equations are similar to those in Eq. (\ref{eq:timelike_geodesics}). As we shall see through specific examples in the 
next section, the other two equations for $\dot{r}$ and $\dot{\theta}$ in this case can be combined to write
\begin{equation}
\frac{\dot{\theta}}{\sqrt{\Theta(\theta)}}=\pm \frac{\dot{r}}{\sqrt{\mathcal{R}(r)}},
\label{eq:r-th}
\end{equation}
where $\Theta(\theta)$ and $\mathcal{R}(r)$ are functions of $\theta$ and $r$ respectively. Integrating the above equation once from the observer's position $(r_o,\theta_o)$ 
to the emission point $(r_e,\pi/2)$ on the accretion disk, we obtain
\begin{equation}
\fint_{\theta_o}^{\pi/2} \frac{d\theta}{\sqrt{\Theta(\theta)}}=\pm \fint_{r_o}^{r_e} \frac{dr}{\sqrt{\mathcal{R}(r)}},
\label{eq:R-TH}
\end{equation}
where the slash notation $\fint$ indicates that these integrals have to be evaluated along the geodesic, while taking 
into account all the turning points in the radial or in the polar motion occurring whenever the corresponding potential 
$\mathcal{R}(r)$ or $\Theta(\theta)$ vanishes. The turning points in $\Theta(\theta)$ are given by $\Theta(\theta)=0$. 
Generally $\Theta(\pi-\theta)=\Theta(\theta)$ (see \cite{shaikh_2019b} for the spherically symmetric case). This means that 
$\Theta(\theta)$ has two turning points, namely $\theta_{tp1}$ and $\theta_{tp2}$, located symmetrically about 
the equatorial plane such that $\theta_{tp1}+\theta_{tp2}=\pi$. Let $0\leq \theta_{tp1}\leq \pi/2$ and 
$\pi/2\leq \theta_{tp2}\leq \pi$, i.e., $\theta_{tp1}$ and $\theta_{tp2}$ lie, respectively, in the upper and 
lower half about the equatorial plane. Also we put the observer above the equatorial plane such that $0\leq \theta_o\leq \pi/2$.

For photons which do not take turn in $\theta$, the left hand side of Eq. (\ref{eq:R-TH}) becomes
\begin{equation}
\fint_{\theta_o}^{\pi/2} \frac{d\theta}{\sqrt{\Theta(\theta)}}=\int_{\theta_o}^{\pi/2} \frac{d\theta}{\sqrt{\Theta(\theta)}}
\label{eq:TH-1}
\end{equation}
However, it is possible that photons, under strong gravitational lensing, may undergo multiple windings around the central 
object and thus may encounter multiple turnings at $\theta_{tp1}$ and $\theta_{tp2}$. Note that 
we integrate the null geodesics backward in time from the observer's position to the emitting point that lies on the disk.
That is, we shoot photons from the observer towards the central object with varying impact parameters.
So, depending on the initial direction of a photon (i.e depending on the sign on the right hand side of 
Eq. (\ref{eq:R-TH}) at the position of the observer), the photon will first encounter one of the turning points. Say
we first choose the `$+$' sign in Eq. (\ref{eq:R-TH}), which corresponds to a photon with $d\theta/dr>0$ at the initial
time, at the observer's position. Then, as the radial coordinate decreases initially along the null geodesic which we shoot from 
the observer, the coordinate $\theta$ must initially decrease. Now if the photon does encounter a turning 
point in $\theta$, it will do so first at $\theta_{tp1}$ and then at $\theta_{tp2}$. After this, 
it is possible to repeat the process and hence the photon might
undergo multiple turnings at both turning points, which will depend on the deflection. Also, the photon might
hit the emitting point on the disk after encountering its last turn at either of the turning points $\theta_{tp1}$ 
and $\theta_{tp2}$. In either of the cases, the left hand side of Eq. (\ref{eq:R-TH}) can be written as (see Eq. (4.11) of \cite{shaikh_2019b}
for details)
\begin{equation}
\fint_{\theta_o}^{\pi/2} \frac{d\theta}{\sqrt{\Theta(\theta)}}=\int_{\theta_o}^{\theta_{tp1}} \frac{d\theta}
{\sqrt{\Theta(\theta)}}-n\int_{\theta_{tp1}}^{\theta_{tp2}} \frac{d\theta}{\sqrt{\Theta(\theta)}}-
\int_{\theta_{tp1}}^{\pi/2} \frac{d\theta}{\sqrt{\Theta(\theta)}},
\label{eq:TH-2}
\end{equation}
where $n=0,1,2,3,\cdots$. For later use, we rewrite the above equation. For that, we use the following
\begin{equation}
\int_{\theta_{tp1}}^{\theta_{tp2}} \frac{d\theta}{\sqrt{\Theta(\theta)}}=\int_{\theta_{tp1}}^{\theta_{o}} \frac{d\theta}
{\sqrt{\Theta(\theta)}}+\int_{\theta_{o}}^{\pi/2} \frac{d\theta}{\sqrt{\Theta(\theta)}}+\int_{\pi/2}^{\theta_{tp2}} 
\frac{d\theta}{\sqrt{\Theta(\theta)}},
\label{eq:ID-1}
\end{equation}
\begin{equation}
\int_{\theta_{tp1}}^{\pi/2} \frac{d\theta}{\sqrt{\Theta(\theta)}}=-\int_{\theta_{tp2}}^{\pi/2} \frac{d\theta}
{\sqrt{\Theta(\theta)}}=\int_{\pi/2}^{\theta_{tp2}} \frac{d\theta}{\sqrt{\Theta(\theta)}},
\label{eq:ID-2}
\end{equation}
where, in Eq. (\ref{eq:ID-2}), we have used the transformation $\theta\rightarrow \pi-\theta$ and the fact 
that $\theta_{tp1}+\theta_{tp2}=\pi$ and $\Theta(\pi-\theta)=\Theta(\theta)$. Equations (\ref{eq:ID-1}) and (\ref{eq:ID-2}) 
can now be combined to obtain
\begin{equation}
\int_{\theta_{o}}^{\theta_{tp1}} \frac{d\theta}{\sqrt{\Theta(\theta)}}-\int_{\theta_{tp1}}^{\pi/2} 
\frac{d\theta}{\sqrt{\Theta(\theta)}}=\int_{\theta_{o}}^{\pi/2} \frac{d\theta}{\sqrt{\Theta(\theta)}}-
\int_{\theta_{tp1}}^{\theta_{tp2}} \frac{d\theta}{\sqrt{\Theta(\theta)}}.
\label{eq:ID-3}
\end{equation}
Using the last equation, Eq. (\ref{eq:TH-2}) can be rewritten as
\begin{equation}
\fint_{\theta_o}^{\pi/2} \frac{d\theta}{\sqrt{\Theta(\theta)}}=\int_{\theta_o}^{\pi/2} \frac{d\theta}
{\sqrt{\Theta(\theta)}}-(n+1)\int_{\theta_{tp1}}^{\theta_{tp2}} \frac{d\theta}{\sqrt{\Theta(\theta)}}.
\label{eq:TH-3}
\end{equation}

Similarly, we now choose the `$-$' sign in Eq. (\ref{eq:R-TH}), i.e., we choose a photon with $d\theta/dr<0$ initially 
at the observer's position. Therefore, since the radial coordinate decreases initially along the photon geodesic 
which we shoot from the observer's position, the coordinate $\theta$ has to increase initially. Then, if the photon 
encounters any turning point in $\theta$, it does so at $\theta_{tp2}$ first and then at $\theta_{tp1}$. After that, 
it may repeat the same and undergo multiple turnings at both the turning points. Following the same procedure 
discussed above, for such a photon, we obtain (see Eq. (4.12) of \cite{shaikh_2019b} for details)
\begin{equation}
\fint_{\theta_o}^{\pi/2} \frac{d\theta}{\sqrt{\Theta(\theta)}}=\int_{\theta_o}^{\theta_{tp2}} 
\frac{d\theta}{\sqrt{\Theta(\theta)}}-m\int_{\theta_{tp2}}^{\theta_{tp1}} \frac{d\theta}
{\sqrt{\Theta(\theta)}}-\int_{\theta_{tp2}}^{\pi/2} \frac{d\theta}{\sqrt{\Theta(\theta)}},
\label{eq:TH-4}
\end{equation}
where $m=0,1,2,3,\cdots$. Now to rewrite the above equation, we can obtain some identities similar 
to those of Eqs. (\ref{eq:ID-1}-\ref{eq:ID-3}) but with the interchange between $\theta_{tp1}$ and $\theta_{tp2}$. 
After doing so, we obtain
\begin{equation}
\int_{\theta_{o}}^{\theta_{tp2}} \frac{d\theta}{\sqrt{\Theta(\theta)}}-\int_{\theta_{tp2}}^{\pi/2} 
\frac{d\theta}{\sqrt{\Theta(\theta)}}=\int_{\theta_{o}}^{\pi/2} \frac{d\theta}{\sqrt{\Theta(\theta)}}-
\int_{\theta_{tp2}}^{\theta_{tp1}} \frac{d\theta}{\sqrt{\Theta(\theta)}}.
\label{eq:ID-4}
\end{equation}
Using the last equation, Eq. (\ref{eq:TH-4}) can be rewritten as
\begin{equation}
\fint_{\theta_o}^{\pi/2} \frac{d\theta}{\sqrt{\Theta(\theta)}}=\int_{\theta_o}^{\pi/2} 
\frac{d\theta}{\sqrt{\Theta(\theta)}}+(m+1)\int_{\theta_{tp1}}^{\theta_{tp2}} \frac{d\theta}{\sqrt{\Theta(\theta)}}.
\label{eq:TH-5}
\end{equation}

Before considering the radial integral in Eq. (\ref{eq:R-TH}), let us discuss the celestial coordinates $\alpha$ and $\beta$ 
which are used to obtain the apparent shape of an image and lie in the celestial plane perpendicular to the line joining 
the observer and the center of the spacetime geometry. The coordinates $\alpha$ and $\beta$ are defined by \citep{celestial}
\begin{equation}
\alpha=\lim_{r_o\to\infty}\left(-r^2\sin\theta\frac{d\phi}{dr}\Big\vert_{(r_o,\theta_o)}\right)~,~~
\beta=\lim_{r_o\to\infty}\left(r^2\frac{d\theta}{dr}\Big\vert_{(r_o,\theta_o)}\right)~,
\label{eq:celestial}
\end{equation}
where $(r_o,\theta_o)$ are the position coordinates of a distant observer. Therefore, in a given spacetime geometry, 
once we obtain the separated geodesic equations yielding $\dot{r}$, $\dot{\theta}$ and $\dot{\phi}$, and put them in the 
above expressions and take the limits, we obtain the expressions for $\alpha$ and $\beta$. Note that the sign of the 
$\beta$ coordinate is the same as $d\theta/dr$, i.e., the same as the sign on the right hand side of Eq. (\ref{eq:R-TH}). 
Therefore, using Eqs. (\ref{eq:TH-1}), (\ref{eq:TH-3}) and (\ref{eq:TH-5}), Eq. (\ref{eq:R-TH}) can be rewritten as
\begin{equation}
\fint_{r_o}^{r_e} \frac{dr}{\sqrt{\mathcal{R}(r)/E^2}}=-\left[k\int_{\theta_{tp1}}^{\theta_{tp2}} \frac{d\theta}
{\sqrt{\Theta(\theta)/E^2}}-sign(\beta)\int_{\theta_o}^{\pi/2} \frac{d\theta}{\sqrt{\Theta(\theta)/E^2}}\right],
\label{eq10mod}
\end{equation}
where $sign(\beta)$ is the sign of the $\beta$ coordinate, and $n$ and $m$ are replaced by a common integer 
$k = 0,1,2,3,\cdots$, and $k=0$ corresponds to Eq. (\ref{eq:TH-1}). In Eq. (\ref{eq10mod}), we have divided 
$\mathcal{R}(r)$ and $\Theta(\theta)$ by $E^2$ because, as we shall see, they contain the energy $E$, 
angular momentum $L$ and the Carter constant $\mathcal{K}$ in general, and when divided by $E^2$, 
these three constants effectively get reduced to two constants $\xi$ and $\eta$ defined by $\xi=L/E$ and 
$\eta=\mathcal{K}/E^2$. $\xi$ and $\eta$ are known as impact parameters. One can check that, for a 
spherically symmetric spacetime, the above equation reduces to Eq. (4.20) of \cite{shaikh_2019b}.

We now consider the radial integral. If there are turning points in $r$, let the outermost one is given by $r_{tp}$ 
such that $\mathcal{R}(r_{tp})=0$. If the photon does not encounter any turning point $r_{tp}$ before it hits the emitting 
point on the disk, then the radial integral in Eq. (\ref{eq:R-TH}) can be written as
\begin{equation}
\fint_{r_o}^{r_e} \frac{dr}{\sqrt{\mathcal{R}(r)/E^2}}=\int_{r_o}^{r_e} \frac{dr}{\sqrt{\mathcal{R}(r)/E^2}}.
\label{eq:R-1}
\end{equation}
However, if it hits the disk after encountering the turning point, then we have
\begin{equation}
\fint_{r_o}^{r_e} \frac{dr}{\sqrt{\mathcal{R}(r)/E^2}}=\int_{r_o}^{r_{tp}} \frac{dr}{\sqrt{\mathcal{R}(r)/E^2}}-\int_{r_{tp}}^{r_e} \frac{dr}{\sqrt{\mathcal{R}(r)/E^2}}.
\label{eq:R-2}
\end{equation}

We now discuss the procedure to find out the solution $r_e$ of Eq. (\ref{eq10mod}) such that $r_e$ lies within the inner and outer boundaries of an accretion disk. The steps are as follows:
\begin{itemize}
\item[i.] We first set the observer's position $(r_o,\theta_o)$ and the coordinates $(\alpha,\beta)$ in the observer's sky. 
This fixes the impact parameters $\xi$ and $\eta$ (or $L$ and $\mathcal{K}$).
\item[ii.] We then increase $k$ from $0$ to $k_{max}$ in steps of $1$. For a given $k$, we use Eq. (\ref{eq:R-1}) in Eq. (\ref{eq10mod}) and solve for $r_e$ such that it lies within the inner and outer boundaries of a disk. For a given $k$, if a real root $r_e$ which lies within the disk boundaries is found, then stop there, note down the root $r_e$, go to step i and change the point $(\alpha,\beta)$ in the observer's sky.
\item[iii.] However, if no real $r_e$ which lies within the disk boundaries is found in step ii for the entire $k$ from $0$ to $k_{max}$, then, this time, we use Eq. (\ref{eq:R-2}) [instead of Eq. (\ref{eq:R-1}) used in step ii] in Eq. (\ref{eq10mod}) and solve for $r_e$ by increasing $k$ from $0$ to $k_{max}$ in steps of $1$. For a given $k$, if a real root $r_e$ which lies within the disk boundaries is found, then stop there, note down the root $r_e$, go to step i and change the point $(\alpha,\beta)$ in the observer's sky. Note that we perform this step only if no root $r_e$ is found in step ii.
\end{itemize}
In our calculation, we take $k_{max}=10$.

The radial integrations given in Eqs. (\ref{eq:R-1}) and (\ref{eq:R-2}) are valid for accretion disks around black holes, 
naked singularities as well as other compact objects. This is valid for a wormhole also, when the disk and the distant 
observer are on the same side of the wormhole throat. However, for wormholes, another scenario can happen, namely that 
the observer and the disk can be on the two opposite sides of the throat. 
In this case, photons which have impact parameters 
less than some critical value do not undergo any turning. Instead, such photons pass through the throat and escape 
to the opposite side \citep{shaikh_2019a}. Therefore, if there is an accretion disk on the side opposite to that of the 
observer, then some of the photons which cross the throat and come to the observer's side will reach the observer 
and produce images. If the wormhole spacetime is written in the radial coordinate with $r_{th}$ being the location of 
the throat, then the radial part in this case can be written as
\begin{equation}
\fint_{r_o}^{r_e} \frac{dr}{\sqrt{\mathcal{R}(r)/E^2}}=\int_{r_o}^{r_{th}} \frac{dr}{\sqrt{\mathcal{R}(r)/E^2}}
-\int_{r_{th}}^{r_e} \frac{dr}{\sqrt{\mathcal{R}(r)/E^2}},
\label{eq:R-3}
\end{equation}
where we have used `$-$' sign before the second term on the right hand side because an ingoing photon becomes 
outgoing after it crosses and leaves the throat. In this scenario, we consider only photons with those impact parameters 
which do not have any turning point outside the throat. For that, we check for the turning point $R(r_{tp})=0$ for a given 
set of impact parameters and consider those impact parameters for which a real solution $r_{tp}$ with $r_{tp}>r_{th}$ 
is not possible. In this case, for the radial part in Eq. (\ref{eq10mod}), we only consider Eq. (\ref{eq:R-3}).

The photon flux as detected by a distant observer is given by \citep{bambi_2012}
\begin{equation}
F_{obs}(r)=g^4\mathcal{F}(r),\quad g = \frac{k_\mu u^\mu_{o}}{k_\nu u^\nu_{e}}=\frac{\sqrt{-g_{tt}
-2g_{t\phi}\Omega-g_{\phi\phi}\Omega^2}}{1-\xi\Omega},
\end{equation}
where $\mathcal{F}(r)$ is the flux obtained in the previous section, $g$ is the redshift factor, $u^\mu_{o} = (1,0,0,0)$ 
is the four-velocity of the distant observer (who is at infinity), $u^\mu_{e} = (\dot{t},0,\dot{\phi},0)$ is the four-velocity 
of the timelike geodesic at the emitting point on the accretion disk, and $k^\mu$ is the four-velocity of the photons. Therefore, once we find $r_e$ for a given set of values of $(\alpha,\beta)$, we assign 
to it the above observed redshifted flux value by putting $r=r_e$.

\subsection{Numerical ray-tracing using Runge-Kutta algorithm}

Note that one cannot use the semi-analytic ray-tracing discussed above if the second order geodesic equations 
for $r$ and $\theta$, i.e., equations for $\ddot{r}$ and $\ddot{\theta}$ can not be integrated once using the Carter 
constant $\mathcal{K}$ to obtain first order equations yielding $\dot{r}$ and $\dot{\theta}$. 
In such a case, we integrate the second order geodesic equations for $r$ and $\theta$ directly 
using the fourth-order Runge-Kutta (R-K) method. However, as we discuss later in this section, in this 
case also, we consider Eq. (\ref{eq:t-phi}) as the first order geodesic equations for $t$ and $\phi$ obtained using the 
two constants of motion $E$ and $L$.

Before discussing the numerical scheme in this section, we first note that numerical solving procedure of the geodesics will involve spacetimes on both the sides of the throat in case of photons crossing the throat, especially in the case when the disk and the observer are on the two opposite sides of the throat. Moreover, the location of the throat corresponds to a coordinate singularity as $g_{rr}$ diverges or $g_{rr}^{-1}$ vanishes there. This may be problematic in the numerical procedure. For these reasons, as mentioned in Sec. \ref{sec:spacetimes}, we write down the metric either in the proper radial coordinate $l$ or in the coordinate $x$. It turns out that it is convenient to write the metric in $x$-coordinate as it may not always be possible to write the same in the proper radial coordinate $l$ \citep{shaikh_2019a}. Now to discuss how to define the coordinate $x$, we first note that $g_{rr}=f(r,\theta)\times r^2/[(r-r_{th})(r+r_{th})]$, where $f$ is some function of $r$ and $\theta$ which is nonzero and finite at the throat $r=r_{th}$. For example, for the Kerr-like wormhole
\begin{equation}
g_{rr}=\frac{\Sigma}{(r-r_{th})(r-r_0)}=\frac{f(r,\theta)r^2}{(r-r_{th})(r+r_{th})},
\end{equation}
where $r_0=M(1+\lambda^2)-\sqrt{M^2(1+\lambda^2)^2-a^2}$ and $f=(r+r_{th})\Sigma/[r^2(r-r_0)]$. Note that $f(r,\theta)$ is nonzero and finite at the throat $r=r_{th}$. Therefore, we define the $x$-coordinate as
\begin{equation}
dx=\pm \frac{rdr}{\sqrt{r^2-r_{th}^2}} \quad \Rightarrow x=\pm \sqrt{r^2-r_{th}^2} \quad \text{or} \quad r^2=x^2+r_{th}^2,
\end{equation}
so that the metric does not have any coordinate singularity as $g_{rr}dr^2\equiv f(x,\theta)dx^2$ does not have any. The $x$-coordinate can now cover both sides of the throat with $x=0$, $x>0$ and $x<0$, respectively, representing the throat, the observer side and the other side. For convenience, we use the radial coordinate $r$ in the discussion below. However, the numerical calculations are done by converting the spacetime metric and the different parameters of the accretion disk in $x$ coordinate.

We now discuss this numerical scheme. 
The basis of a locally inertial frame at the observer's location is chosen 
as $\{ \hat{e}_{(t)},\hat{e}_{(r)},\hat{e}_{(\theta)},\hat{e}_{(\phi)} \}$, which is related to the 
coordinate basis $\{\partial_{t},\partial_{r},\partial_{\theta},\partial_{\phi}\}$ by the following transformation relations.
\begin{equation}
\hat{e}_{(\alpha)}= \hat{e}^{\mu}_{(\alpha)}\partial_{\mu}, ~~~~~ 
\text{where} ~~~~~ g_{\mu\nu}\hat{e}^{\mu}_{(\alpha)} \hat{e}^{\nu}_{(\beta)}=\eta_{(\alpha)(\beta)}
\end{equation}
Here, $\eta_{(\alpha)(\beta)}$ represents the usual Minkowski metric. Subscripts or superscripts 
written within the parenthesis indicate components in locally inertial frame and without parenthesis 
represent components in global coordinates. To be specific, we have considered the Zero Angular Momentum 
Observer (ZAMO) to construct our locally inertial frame. In this ZAMO frame, the above transformation relations become
\begin{equation}
\hat{e}_{(t)}=\zeta\partial_{t} + \gamma\partial_{\phi}, ~~ \hat{e}_{(r)}=A^{r}\partial_{r}, ~~ 
\hat{e}_{(\theta)}=A^{\theta}\partial_{\theta}, \hat{e}_{(\phi)}=A^{\phi}\partial_{\phi}
\end{equation}
Applying the conditions $\hat{e}_{(t)}\cdot\hat{e}_{(t)}=-1$, $\hat{e}_{(t)}\cdot\hat{e}_{(\phi)}=0$, 
$\hat{e}_{(r)}\cdot\hat{e}_{(r)}=1$, $\hat{e}_{(\theta)}\cdot\hat{e}_{(\theta)}=1$ and 
$\hat{e}_{(\phi)}\cdot\hat{e}_{(\phi)}=1$, we obtain
\begin{equation}
\zeta=\sqrt{\frac{g_{\phi\phi}}{g}}, ~~ \gamma=-\frac{g_{t\phi}}{g_{\phi\phi}}\zeta~, ~~ 
A^r=\frac{1}{\sqrt{g_{rr}}}, ~~ A^{\theta}=\frac{1}{\sqrt{g_{\theta\theta}}}~, ~~ A^{\phi}=\frac{1}{\sqrt{g_{\phi\phi}}}
\end{equation}
where $g=g^2_{t\phi} - g_{tt}g_{\phi\phi}$. With respect to the ZAMO frame, the locally measured four-momentum components of a photon are given by the following expressions : 
\begin{align*}
P_{(t)}&=\hat{e}^{\mu}_{(t)}P_{\mu}=(\zeta P_t + \gamma P_{\phi}), ~~~~~ {\rm or}, ~~~~~ P^{(t)}=-(\zeta P_t + \gamma P_{\phi})\\
P_{(r)}&=\hat{e}^{\mu}_{(r)}P_{\mu}=\frac{1}{\sqrt{g_{rr}}}P_r, ~~~~~~~~~~ {\rm or}, ~~~~~ P^{(r)}=\frac{1}{\sqrt{g_{rr}}}P_r, \\
P_{(\theta)}&=\hat{e}^{\mu}_{(\theta)}P_{\mu}=\frac{1}{\sqrt{g_{\theta\theta}}}P_{\theta}, 
~~~~~~~~~~ {\rm or}, ~~~~~ P^{(\theta)}=\frac{1}{\sqrt{g_{\theta\theta}}}P_{\theta}, \\
P_{(\phi)}&=\hat{e}^{\mu}_{(\phi)}P_{\mu}=\frac{1}{\sqrt{g_{\phi\phi}}}P_{\phi}, 
~~~~~~~~~ {\rm or}, ~~~~~ P^{(\phi)}=\frac{1}{\sqrt{g_{\phi\phi}}}P_{\phi}
\end{align*}
Due to time translational and axial symmetry, the corresponding two conserved quantities or constants of 
motion are, $P_t=-E$ and $P_{\phi}=L$ respectively, where $E$ represents the total energy and $L$ stands for 
the angular momentum of a photon with respect to a static observer at infinity. 
Since $P_r=g_{rr}P^r=g_{rr}\dot{r}$ and $P_{\theta}=g_{\theta\theta}P^{\theta}=g_{\theta\theta}\dot{\theta}$ 
(considering $P^r=\dot{r}$ and $P^{\theta}=\dot{\theta}$), 
the components of four momentum of a photon in ZAMO frame will be
\begin{equation}
P^{(t)}=E\zeta - L\gamma, ~~~~ P^{(r)}=\sqrt{g_{rr}} ~ \dot{r}~, ~~~~ 
P^{(\theta)}=\sqrt{g_{\theta\theta}} ~ \dot{\theta}, ~~~~ P^{(\phi)}=\frac{L}{\sqrt{g_{\phi\phi}}}	\label{four_mom}
\end{equation}
where an `overdot' as usual means a derivative with respect to the affine parameter along the geodesics.

Let us consider that the observer is located at $r=r_o$, $\theta=\theta_o$ and $\phi=\phi_o=0$. Then the position coordinates of an image in the observer’s sky can be obtained as \citep{Wang1} (see also \citep{Johannsen,Cunha1})
\begin{equation}
x_o=-r_o \left. \frac{P^{(\phi)}}{P^{(r)}}\right\rvert_{(r_o,\theta_o)}=
-\left. \frac{r_o L}{\sqrt{g_{rr}}\sqrt{g_{\phi\phi}}~\dot{r}}\right\rvert_{(r_o,\theta_o)}, \label{celestial_x}
\end{equation}
\begin{equation}
y_o=r_o \left. \frac{P^{(\theta)}}{P^{(r)}}\right\rvert_{(r_o,\theta_o)}=
\left. \frac{r_o\sqrt{g_{\theta\theta}}~\dot{\theta}}{\sqrt{g_{rr}}~\dot{r}}\right\rvert_{(r_o,\theta_o)} \label{celestial_y}
\end{equation}
Note that the celestial coordinates in Eq. (\ref{eq:celestial}) are defined for an asymptotic observer, whereas 
the coordinates $x_o$ and $y_o$ above are defined for the arbitrary observer position $r_o$. It is 
straightforward to show that both these sets of coordinates are equivalent to each other in the limit $r_o\to \infty$. 
From the normalization of four momentum of a photon, we get
\begin{equation}
	(P^{(t)})^2=(P^{(r)})^2 + (P^{(\theta)})^2 + (P^{(\phi)})^2 \label{mom_norm}
\end{equation}
Using Eqs. (\ref{celestial_x}) and (\ref{celestial_y}) in Eq. (\ref{mom_norm}), we obtain
\begin{equation}
P^{(r)}=\left. \pm\frac{P^{(t)}~r_o}{\sqrt{r_o^2 + x_o^2 + y_o^2}}\right\rvert_{(r_o,\theta_o)} \implies ~~~ 
\dot{r}=\left. \pm\frac{1}{\sqrt{g_{rr}}}\frac{P^{(t)}~r_o}{\sqrt{r_o^2 + x_o^2 + y_o^2}}\right\rvert_{(r_o,\theta_o)} \label{rdot}
\end{equation}
Now, using Eqs. (\ref{celestial_y}) and (\ref{rdot}), we get
\begin{equation}
\dot{\theta}=\left. \pm\frac{1}{\sqrt{g_{\theta\theta}}}\frac{P^{(t)}~y_o}
{\sqrt{r_o^2 + x_o^2 + y_o^2}}\right\rvert_{(r_o,\theta_o)} \label{thdot}
\end{equation}
Again, combining Eqs. (\ref{celestial_x}) and (\ref{rdot}), and from the first expression of Eq. (\ref{four_mom}), 
we find
\begin{equation}
L=\left. \mp\frac{P^{(t)}~x_o~\sqrt{g_{\phi\phi}}}{\sqrt{r_o^2 + x_o^2 + y_o^2}}\right\rvert_{(r_o,\theta_o)}~, ~~~ 
\text{and} ~~~~~~ E=\left.\frac{1}{\zeta}\left(P^{(t)} + L\gamma\right)\right\rvert_{(r_o,\theta_o)} \label{L_E}
\end{equation}
%

From Eqs. (\ref{rdot}), (\ref{thdot}) and (\ref{L_E}), we see that, once the observer's position coordinates, 
i.e., $r_o$ and $\theta_o$ are given, we can determine $L$, $E$, $\dot{r}$ and $\dot{\theta}$ for a given 
set of values of the celestial coordinates $x_o$ and $y_o$ in the observer's sky. The only unknown 
parameter that still remains is $P^{(t)}$. Actually, we can easily do away with this parameter and set it 
equal to unity just by re-normalizing the affine parameter along the null geodesics. Therefore, taking $P^{(t)}=1$, 
the final expressions of $L$, $E$, $\dot{r}$ and $\dot{\theta}$ at the observer's location take the form

\begin{align}
\nonumber L&=\left. \mp\frac{x_o~\sqrt{g_{\phi\phi}}}{\sqrt{r_o^2 + x_o^2 + 
y_o^2}}\right\rvert_{(r_o,\theta_o)}, ~~~~~~~~~~~~~~~ E=\left.\frac{1}{\zeta}
\left(1 + L\gamma\right)\right\rvert_{(r_o,\theta_o)} \\
\dot{r}&=\left. \pm\frac{1}{\sqrt{g_{rr}}}\frac{r_o}{\sqrt{r_o^2 + x_o^2 + y_o^2}}\right\rvert_{(r_o,\theta_o)}, 
~~~~~~~~ \dot{\theta}=\left. \pm\frac{1}{\sqrt{g_{\theta\theta}}}\frac{y_o}{\sqrt{r_o^2 + x_o^2 + 
y_o^2}}\right\rvert_{(r_o,\theta_o)} \label{initial_values}
\end{align}
With the equations of the problem properly constructed, let us now discuss the numerical recipe to be followed to 
solve the null geodesic equations and obtain the shadow images. This consists of three main steps. 
\begin{itemize}
\item
First, we start with the null geodesic equations which are given below.
\begin{equation}
 \dot{t}=\frac{1}{g}\left(g_{t\phi}L + g_{\phi\phi}E\right)~,~~\dot{\phi}=-\frac{1}{g}\left(g_{tt}L + g_{t\phi}E\right)~,~~
\ddot{r}=-\Gamma^r_{\mu\nu}\dot{x}^{\mu}\dot{x}^{\nu}~,~~\ddot{\theta}=-\Gamma^{\theta}_{\mu\nu}\dot{x}^{\mu}\dot{x}^{\nu}
\end{equation}
where $g=g_{t\phi}^2-g_{tt}g_{\phi\phi}$. Since we have two constants of motion ($E$ and $L$) in an arbitrary stationary and axially symmetric spacetime, it is 
convenient to use the first order 
differential equations in $t$ and $\phi$ which are obtained by integrating their corresponding second order 
geodesic equations once. So we have two first order and two second order differential equations, and to solve 
these coupled equations numerically, we need a total of six initial conditions, together with the initial values of 
$E$ and $L$ which remain constant throughout a geodesic.

\item
The initial values of the coordinates at the observer's location are chosen to be, $t_o=0$, $r_o=10^4M$, 
$\theta_o=80^{\circ}$ and $\phi_o=0$. These constitute four of the initial conditions, and the remaining two, i.e., 
$\left.\dot{r}\right\rvert_{(r_o,\theta_o)}$ and $\left.\dot{\theta}\right\rvert_{(r_o,\theta_o)}$, as well as the initial 
values of $E$ and $L$, are given by Eq. (\ref{initial_values}) for a chosen set of values of the coordinates $(x_o,y_o)$. So with all the initial conditions specified, we solve the differential equations 
from observer to the lensing object using a standard fourth order R-K method and obtain a single geodesic 
trajectory of a photon corresponding to the chosen initial values of $(x_o,y_o)$. If we take a different set 
of values of $(x_o,y_o)$, keeping $(t_o,r_o,\theta_o,\phi_o)$ fixed, i.e., the observer's position fixed, 
it will yield a different trajectory.

\item
If, for a specific set of values of the coordinates $(x_o,y_o)$, the corresponding photon geodesic, 
originating from the observer, hits the accretion disk on the equatorial plane, we note down the corresponding 
hitting point $r_e$, calculate the redshifted flux $F_{obs}(r)$ by putting $r=r_e$ and assign this flux to the coordinates $(x_o,y_o)$. On the other hand, if, for some other specific value of $(x_o,y_o)$, 
the corresponding photon orbit swirls in the lensing object and doesn't reach to any point on the disk, we 
consider that point of the observer's sky a dark spot and set a zero flux to that $(x_o,y_o)$. We, thus, 
vary the values of $(x_o,y_o)$ to cover the observer's sky, and depending on the final fate of the 
geodesics as described, determine the flux of each $(x_o,y_o)$ point. In this way, we obtain the 
intensity map of the images of accretion disks in the observer's sky for a fixed position of the observer.
\end{itemize}

This concludes our discussion of the numerical recipes used in this paper, and we are now 
ready to present our main results. 

\section{Images of wormholes with accretion disks}
\label{sec:images}

To produce the intensity map of the images, we have used both the techniques discussed in section \ref{sec:tracing} 
and have found that the images obtained using both the techniques match each other. In order to use the 
semi-analytic ray tracing technique, we need to obtain the first order separated geodesic equations. 
For the Teo wormhole given in Eq. (\ref{eq:teo_wormhole}), these are given by \citep{nedkova_2013}
\begin{equation}
\dot{t}=\frac{E-\omega L}{N^2}~,~ \dot{\phi}=\frac{L}{r^2K^2\sin^2\theta}+\frac{\omega(E-\omega L)}{N^2}~,~
\frac{N}{\left(1-\frac{b}{r}\right)^{1/2}}\frac{dr}{d\lambda} = \pm\sqrt{R(r)}~,~
r^2K^2\frac{d\theta}{d\lambda} = \pm\sqrt{T(\theta)},
\label{eq:r-th-exp}
\end{equation}
where we have defined
\begin{equation}
R(r)=\left(E-\omega L\right)^2 - (\mathcal{K}+L^2) \frac{N^2}{r^2K^2}~,~
T(\theta) = \mathcal{K} - \frac{\cos^2\theta}{\sin^2\theta}L^2~.
\end{equation}
Note that the constant $Q$ in Ref. \cite{nedkova_2013} is related to $\mathcal{K}$ through $Q=\mathcal{K}+L^2$. Comparing Eq. (\ref{eq:r-th}) with Eq. (\ref{eq:r-th-exp}), we find that $\Theta(\theta)=T(\theta)$ and
\begin{equation}
\mathcal{R}(r)=\frac{r^4K^4}{N^2}\left(1-\frac{b}{r}\right)R(r).
\end{equation}
Note that when we divide $\mathcal{R}(r)$ by $E^2$, as is done in Eq. (\ref{eq10mod}), we have effectively two impact parameters $\xi=L/E$ and $\eta=\mathcal{K}/E^2$, instead of three constants of motion $E$, $L$ and $\mathcal{K}$. For the Teo wormhole, the metric functions we use are given in Eq. (\ref{eq:metric_choice}). In this case, the expressions of the celestial coordinates defined in Eq. (\ref{eq:celestial}) turn out to be
\begin{equation}
\alpha= -\frac{\xi}{\sin\theta_{o}}, \quad
\beta=\pm \left(\eta - \xi^2\cot^2\theta_o \right)^{1/2}.
\end{equation}

For the Kerr-like wormhole given in Eq. (\ref{eq:KerrWH}), the first order separated geodesic equations are given by \citep{amir_2018}
\begin{eqnarray}
\Sigma \dot{t} &=& -a \left(aE \sin^2 \theta - L\right) 
+ \frac{(r^2 + a^2) \mathcal{P}}{r^2 + a^2 - 2 M r}~,~
\Sigma \dot{r}= \pm  \sqrt{\mathcal{R}}~,~ \Sigma \dot{\theta}= \pm  \sqrt{\Theta}~,\nonumber\\
\Sigma \dot{\phi} &=& -\left(aE - L\csc^2 \theta \right) 
+ \frac{a \mathcal{P}}{r^2 + a^2 - 2 M r}
\end{eqnarray}
where we have defined
\begin{eqnarray}\label{quant}
\mathcal{R} &=& \frac{\hat{\Delta}(r)
\left[ \mathcal{P}^2 -\Delta(r) 
\left[\mathcal{K}+ (L_{z}-a E)^2 \right] \right] }{\Delta(r)}~,\quad \Theta = \mathcal{K} +\cos^2 \theta \left(a^2E^2-L_{z}^2\csc^2 \theta \right)~, \nonumber\\
\mathcal{P} &=& (r^2+a^2)E-aL_{z}, \quad \Delta(r)=r^2-2Mr+a^2.
\end{eqnarray}
In this case, $\mathcal{R}(r)$ and $\Theta(\theta)$ in Eq. (\ref{eq:r-th}) are the same as those 
in the above equations. In this case, the expressions of the celestial coordinates defined in Eq. (\ref{eq:celestial}) turn out to be
\begin{equation}
\alpha= -\frac{\xi}{\sin\theta_{o}}, \quad
\beta=\pm \left(\eta +a^2\cos^2\theta_o - \xi^2\cot^2\theta_o \right)^{1/2}.
\end{equation}
Once we know $\mathcal{R}(r)$ and $\Theta(\theta)$ for a given 
wormhole, we use them in the semi-analytic ray tracing technique and produce the images.

\begin{table}[h!]
\centering
\caption{The positions of the wormhole throat ($r_{th}$) or black hole outer horizon ($r_+$) and the inner edge ($r_{in}$) of the disk for different parameter values. Generally, the inner edge $r_{in}$ of the disk is at the marginal stable orbit $r_{ms}$. However, when there is no marginal stable orbit outside the wormhole throat, the throat itself acts as the position of the disk inner edge (see the discussion in Sec. \ref{sec:disks}).}
 \begin{tabular}{| c | c | c | c | c | c | c | c|} 
 \hline\hline
  & \multicolumn{2}{c|}{Teo wormhole} & \multicolumn{3}{c|}{Kerr-like wormhole} & \multicolumn{2}{c|}{Kerr black hole} \\
 $a/M$ & $r_{th}/M$ & $r_{in}/M$ & $\lambda^2$ & $r_{th}/M$ & $r_{in}/M$ & $r_{+}/M$ & $r_{in}/M$ \\
 \hline
  0 & 2 & 2 & 0.1 & 2.2 & 6.0 & 2 & 6 \\
   &  &  & 0.8 & 3.6 & 6.0 & & \\
 \hline
  0.7 & 2 & 2 & 0.1 & 1.949 & 3.393 & 1.714 & 3.393 \\
   &  &  & 0.8 & 3.458 & 3.458 &  & \\
 \hline\hline
 \end{tabular}
\label{Table1}
\end{table}

\begin{figure}[ht]
\centering
\subfigure[~$a/M = 0.0$, black hole]{\includegraphics[scale=0.54]{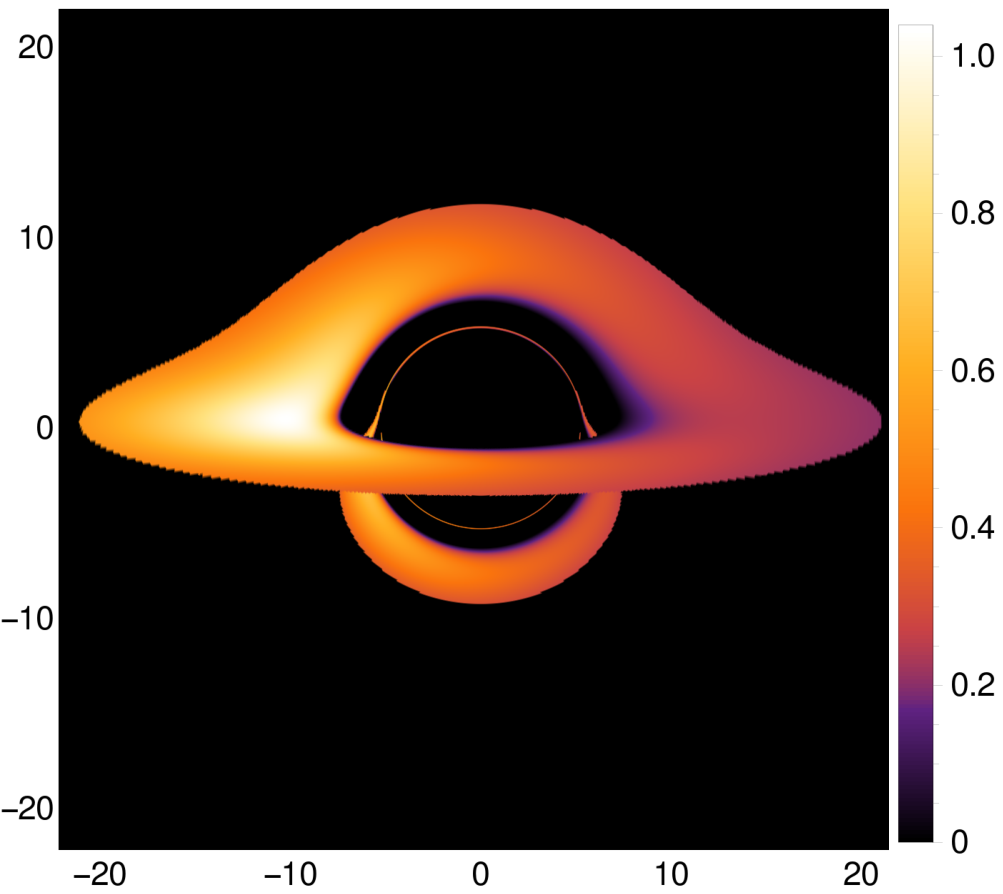}\label{fig:ExpWH_a}}\hspace{0.1cm}
\subfigure[~$a/M = 0.0$, wormhole, same side]{\includegraphics[scale=0.54]{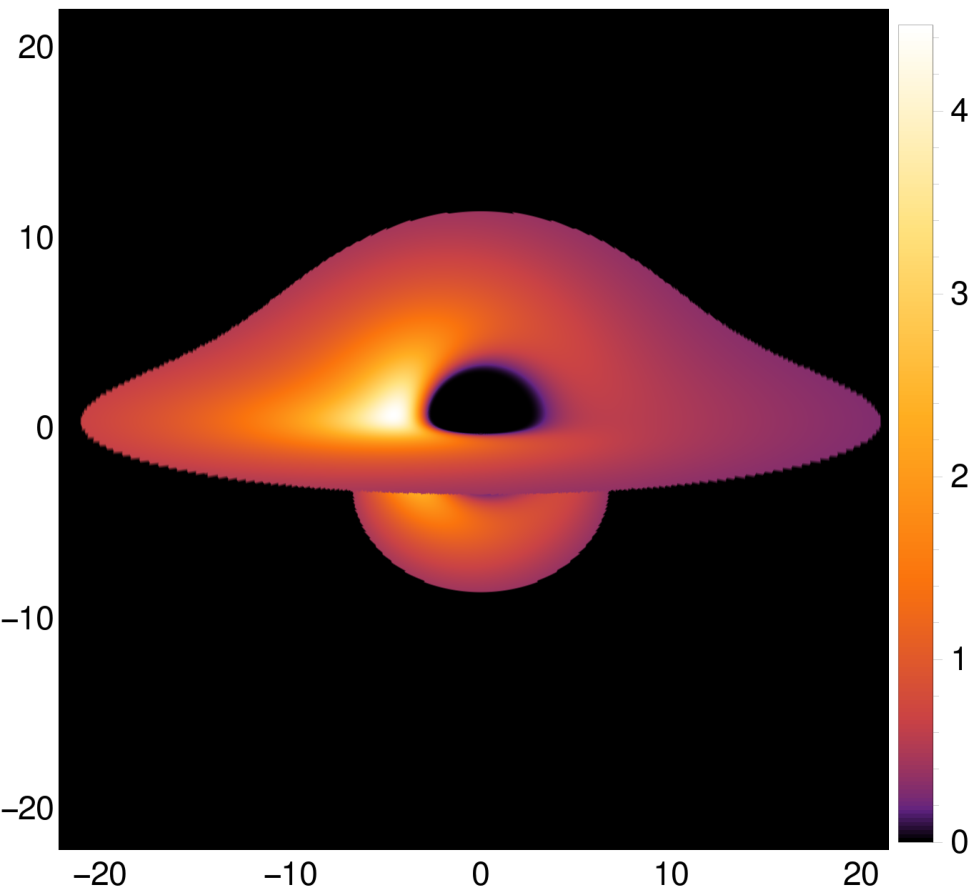}\label{fig:ExpWH_b}}\hspace{0.1cm}
\subfigure[~$a/M = 0.0$, wormhole, other side]{\includegraphics[scale=0.54]{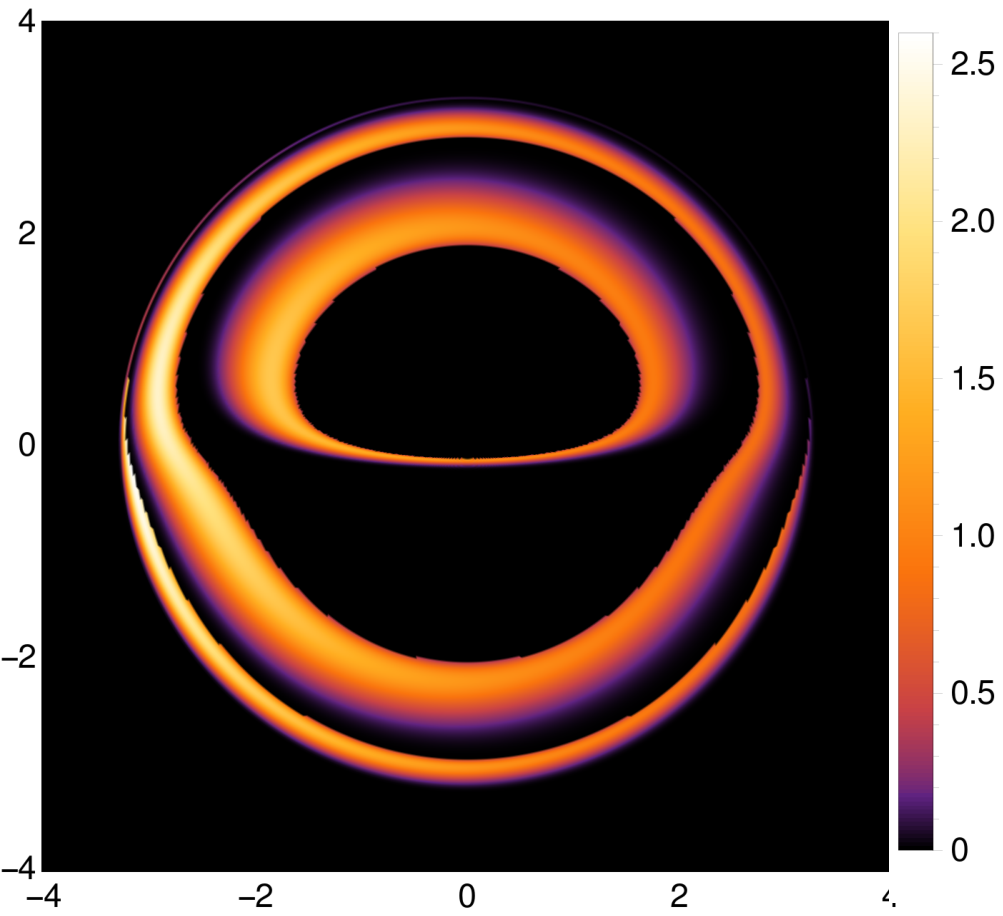}\label{fig:ExpWH_c}}
\subfigure[~$a/M = 0.7$, black hole]{\includegraphics[scale=0.53]{KerrWH_lambda00_a070_SS1.eps}\label{fig:ExpWH_d}}\hspace{0.1cm}
\subfigure[~$a/M = 0.7$, wormhole, same side]{\includegraphics[scale=0.54]{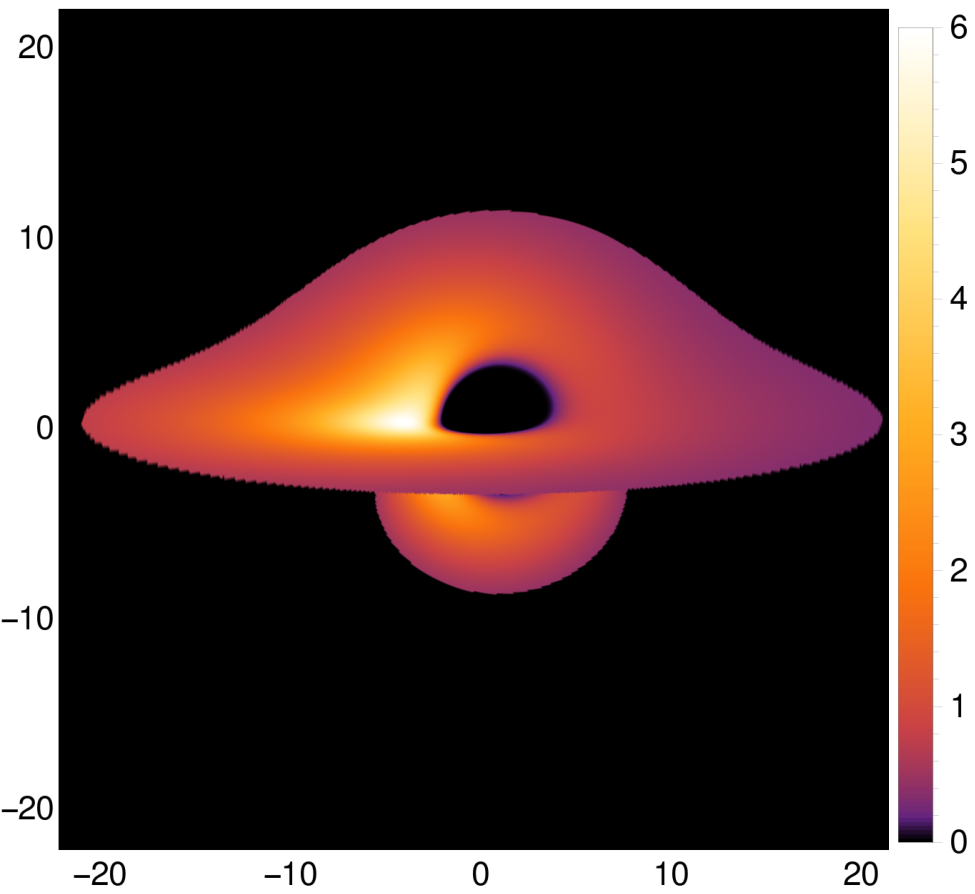}\label{fig:ExpWH_e}}\hspace{0.1cm}
\subfigure[~$a/M = 0.7$, wormhole, other side]{\includegraphics[scale=0.54]{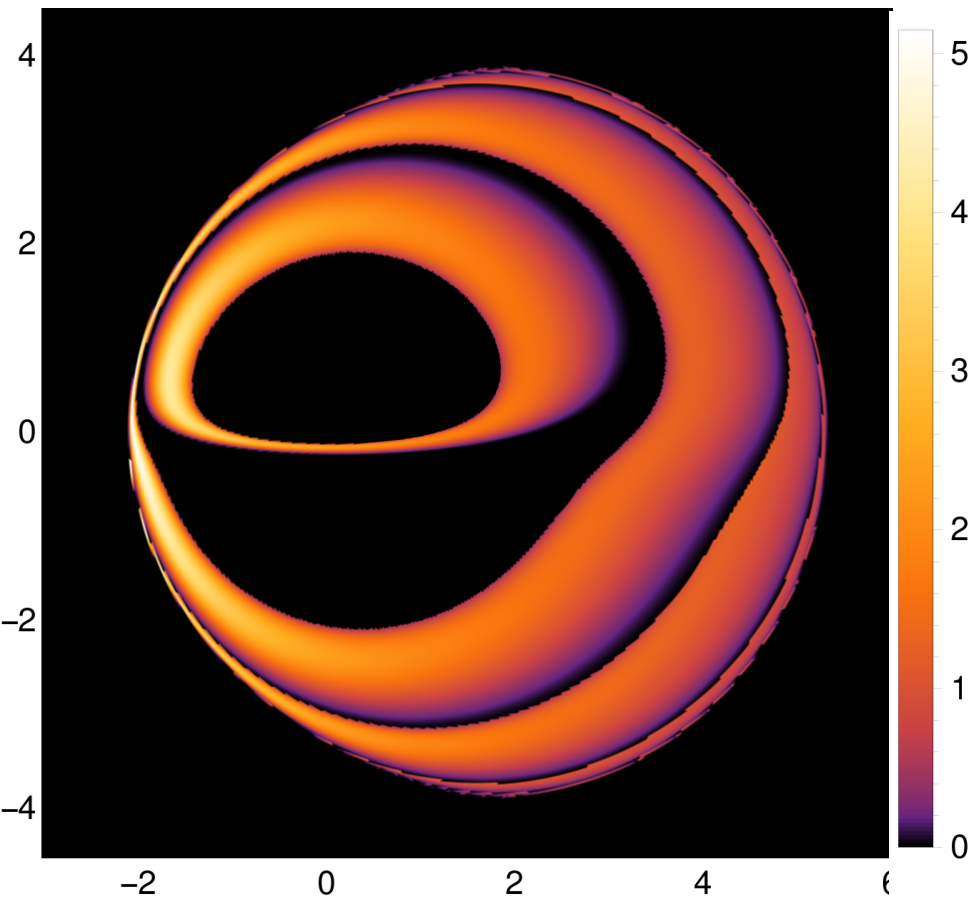}\label{fig:ExpWH_f}}
\caption{The images of a Kerr black hole [(a) and (d)] and a Teo wormhole [(b), (c), (e) and (f)] with accretion disks. (b) and (e) are for the case when the disk is on the observer's side, and (c) and (f) are for the case when it is on the other side. The outer edge of the disk is at $r=20M$, and the position of its inner edge for different parameters is shown in Table \ref{Table1}. The observer's inclination angle is $\theta_{o}=80^{\circ}$. The observer is placed at the radial coordinate $r=10^4M$, which corresponds effectively to the asymptotic infinity. In order to get rid of the parameters $M$ and $\dot{M}$, we have normalized the fluxes by the maximum flux observed for the Kerr black hole with $a=0$. Also, we have plotted the square-root of the normalized flux and rescaled the color function for better looking. The color bars show the values of the square-root of the flux. All spatial coordinates are in units of $M$.}
\label{fig:ExpWH}
\end{figure}

\begin{figure}[ht]
\centering
\subfigure[~$a = 0.0$, $\lambda^2=0.1$, same side]{\includegraphics[scale=0.54]
{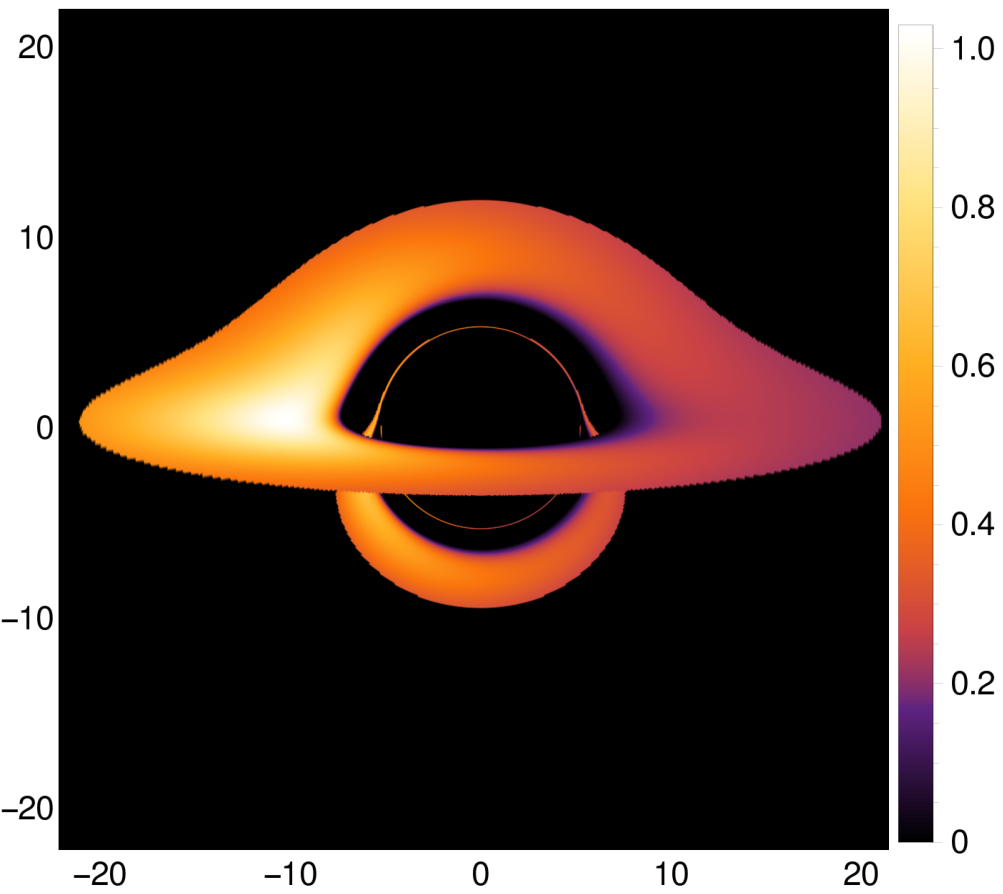}\label{fig:KerrWH_a}}\hspace{0.1cm}
\subfigure[~$a = 0.0$, $\lambda^2=0.1$, other side]{\includegraphics[scale=0.54]
{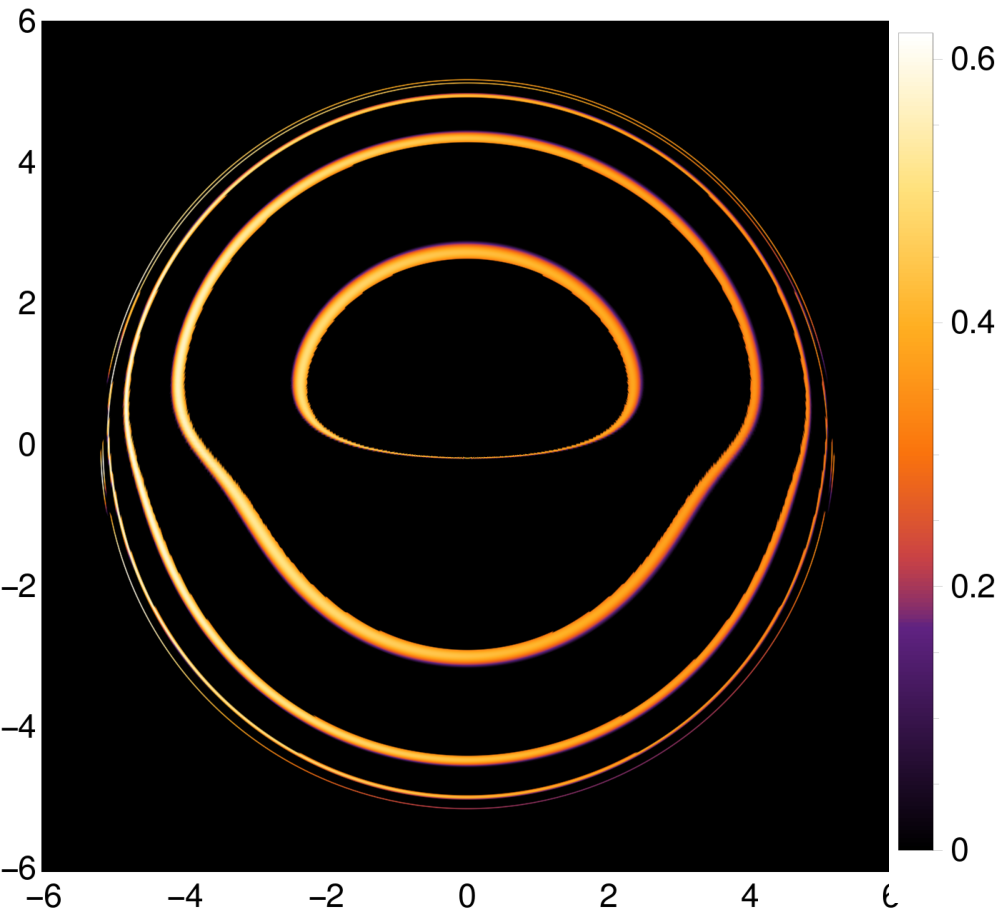}\label{fig:KerrWH_b}}\hspace{0.1cm}
\subfigure[~$a = 0.0$, $\lambda^2=0.8$, other side]{\includegraphics[scale=0.54]{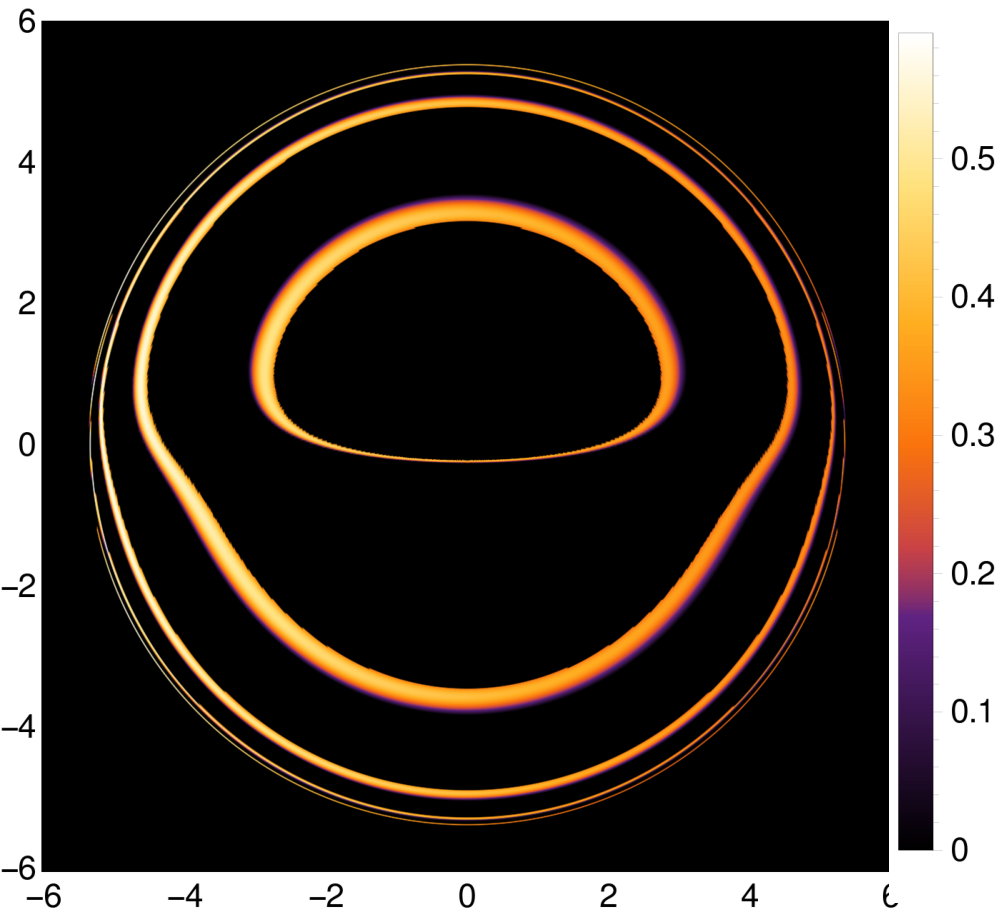}\label{fig:KerrWH_c}}
\subfigure[~$a = 0.7$, $\lambda^2=0.1$, same side]{\includegraphics[scale=0.54]
{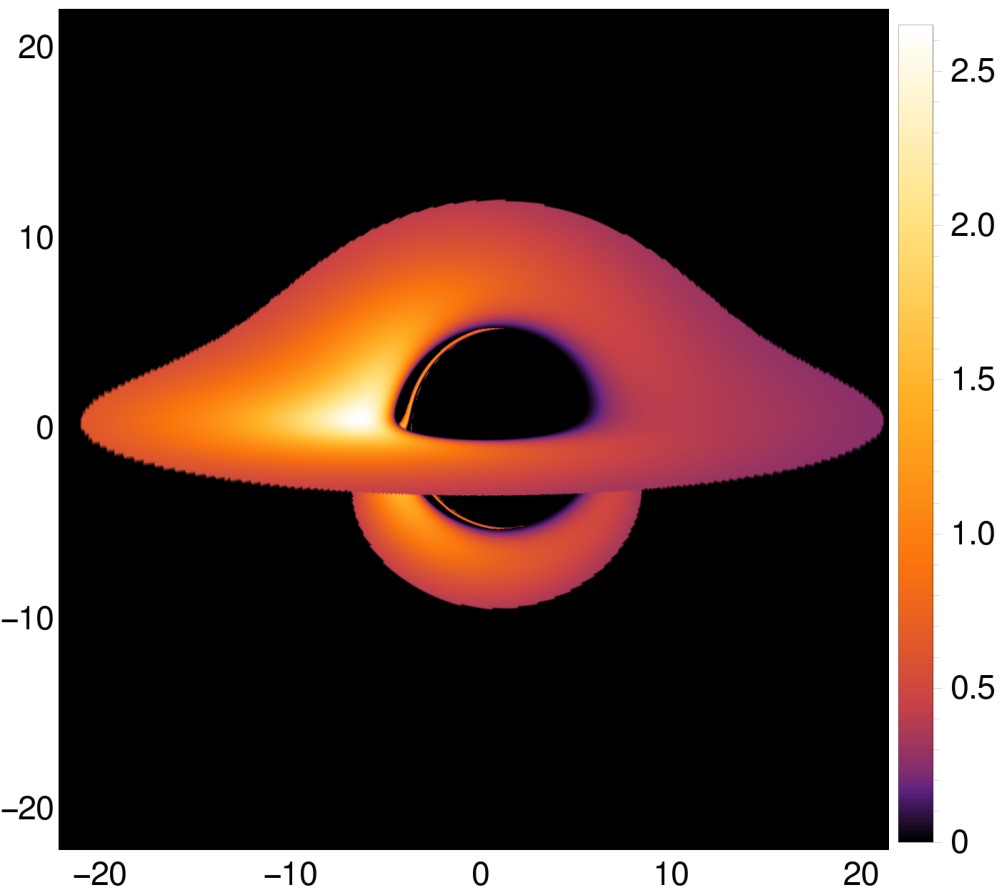}\label{fig:KerrWH_d}}\hspace{0.1cm}
\subfigure[~$a = 0.7$, $\lambda^2=0.1$, other side]{\includegraphics[scale=0.54]{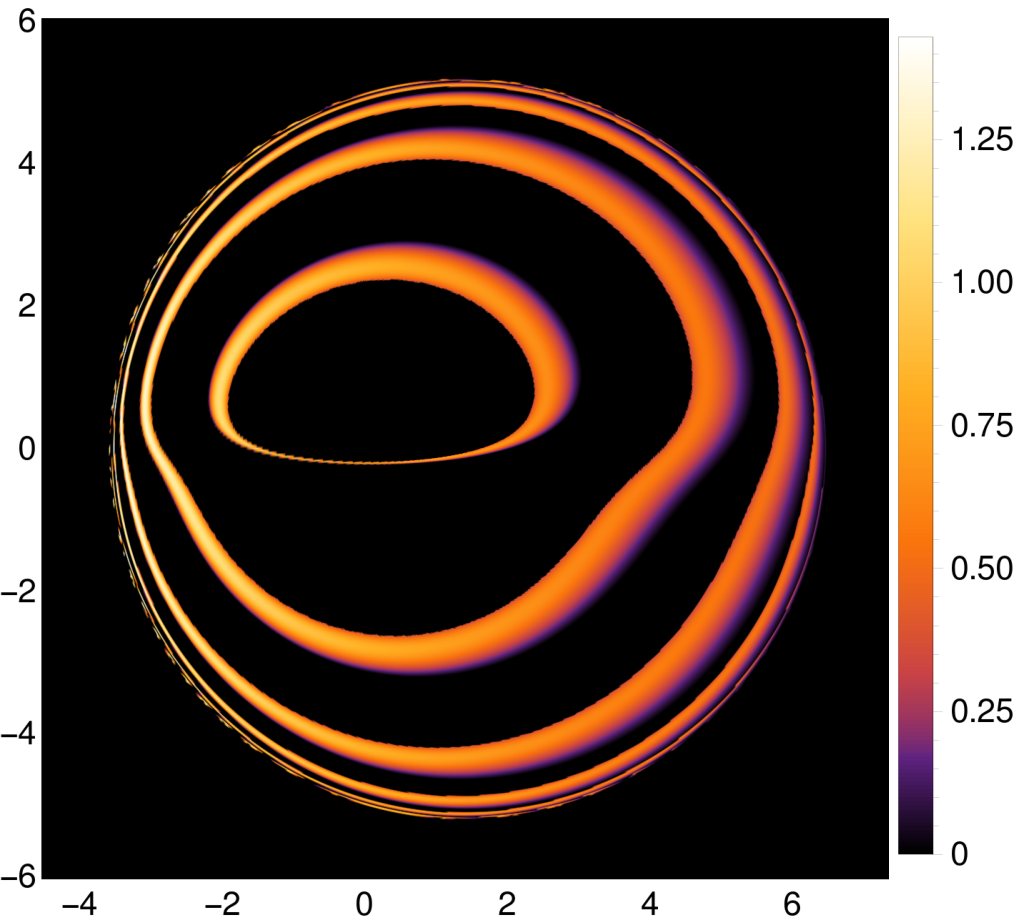}\label{fig:KerrWH_e}}
\subfigure[~$a = 0.7$, $\lambda^2=0.8$, other side]{\includegraphics[scale=0.54]{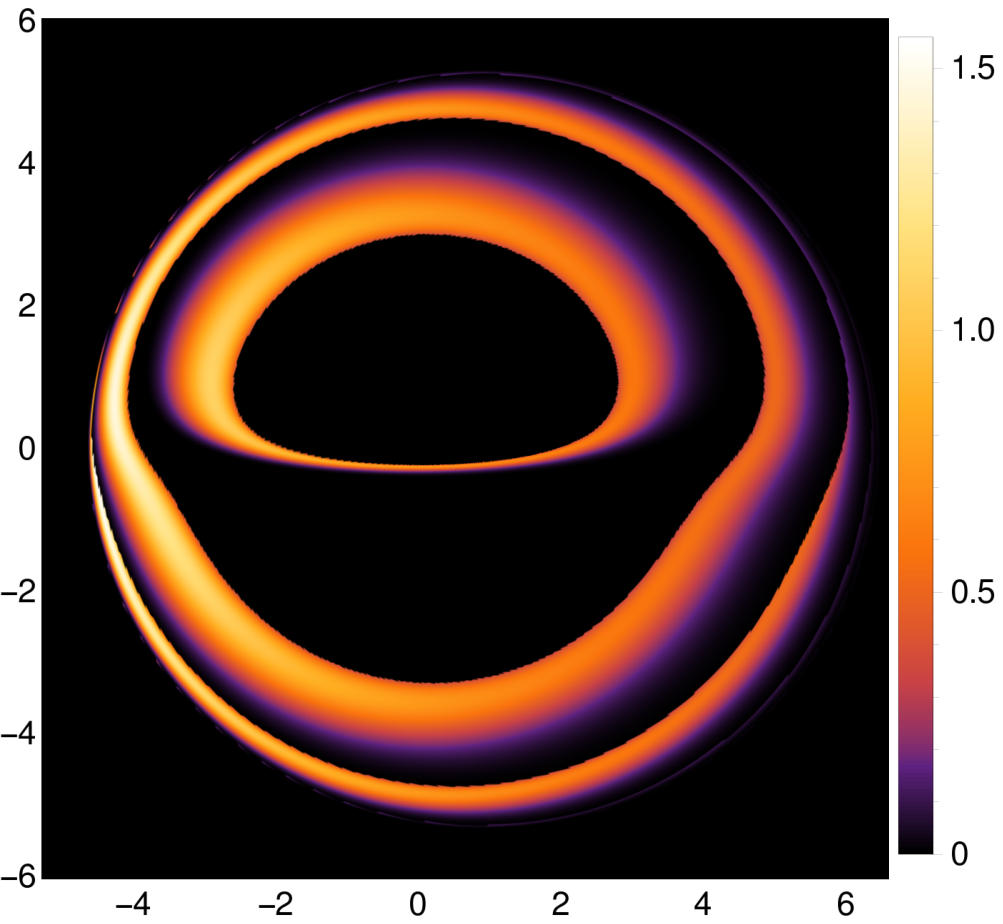}\label{fig:KerrWH_f}}
\caption{The images of a Kerr-like wormhole with an accretion disk when the disk is on the observer's side [(a) and (d)], and 
when it is on the other side [(b), (c), (e) and (f)]. The outer edge of the disk is at $r=20M$, and the position of its inner edge for 
different parameters is shown in Table \ref{Table1}. The observer's inclination angle is $\theta_{o}=80^{\circ}$. The 
observer is placed at the radial coordinate $r=10^4M$, which corresponds effectively to the asymptotic infinity. In 
order to get rid of the parameters $M$ and $\dot{M}$, we have normalized the fluxes by the maximum flux 
observed for the Kerr black hole with $a=0$. Also, we have plotted the square-root of the normalized flux and 
rescaled the color function for better looking. The color bars show the values of the square-root of the flux. 
All spatial coordinates are in units of $M$.}
\label{fig:KerrWH}
\end{figure}

Table \ref{Table1} shows the position of the inner edge of an accretion disk for different parameter values. Figures \ref{fig:ExpWH} and \ref{fig:KerrWH} show the images of the accretion disk around a wormhole 
and a black hole. When the disk is on the same side, the image for a given set of parameters may or may not 
qualitatively differ from or mimic that of a black hole. In this case, the images for the Teo wormhole 
qualitatively differ from those of a black hole as the image of the black hole contains a thin ring image inside the direct image, whereas the image of the wormhole does not [compare between Figs. \ref{fig:ExpWH_a} and \ref{fig:ExpWH_b} and between Figs. \ref{fig:ExpWH_d} and \ref{fig:ExpWH_e}]. This thin ring image is formed by light rays which undergo strong gravitational lensing due to the photon sphere which lies outside the event horizon. For the Teo wormhole, we don't have any photon sphere outside the throat. Although the throat in this case acts as an effective photon sphere, before a photon starts winding around the throat due strong lensing, it hits the disk as the disk extends upto the throat. As a result, we don't have the thin ring image for the Teo wormhole.
However, the images for the Kerr-like wormhole in this scenario mimic those for Kerr black 
hole [compare between Figs. \ref{fig:ExpWH_a} and \ref{fig:KerrWH_a} and between Figs. \ref{fig:ExpWH_d} and \ref{fig:KerrWH_d}].

The most striking differences in the images, however, occur when we consider the disk to be on the 
other side of the throat. The images in this case have unique characteristic features. A single intensity 
map of a disk image contains multiple images [see Figs. \ref{fig:ExpWH_c}, \ref{fig:ExpWH_f}, \ref{fig:KerrWH_b}, 
\ref{fig:KerrWH_c}, \ref{fig:KerrWH_e} and \ref{fig:KerrWH_f}]. This unique characteristic feature of 
the images can in principle distinguish a wormhole background from a black hole. Here it should be noted that, when there are disks on both the sides of the throat, the images will be the superposition of the images of the disk on the observer's side and those of the disk on the other side. One such example is shown in Fig. (\ref{fig:KerrWH_BS}), where we also provide a zoomed-in version to display the
features more prominently. The striking difference with the black hole example of 
Fig.(\ref{fig:ExpWH_d}) is clear. Figs. (\ref{fig:ExpWH_d}) and (\ref{fig:finalA}) are the ones shown in the introduction. 

\section{conclusions}
\label{sec:conclusions}

Currently, there is much interest in observational features of space-times with an event horizon. Images 
from the Event Horizon Telescope will be pivotal in the understanding of such features, which are crucial
to further our knowledge of strong gravity, and ultimately probe the limits of quantum gravity. 
Various works have appeared in the literature regarding the observational distinction between objects with
and without an event horizon, and the broad 
consensus till now is that several observational features of black holes might in fact be indistinguishable from
those in space-times without an event horizon. The fact that wormhole geometries can mimic many 
aspects of black holes has been well recognized since a decade, and spur the interest in quantifying
these in terms of the recently available experimental data. 

\begin{figure}[ht]
\centering
\subfigure[~$a = 0.7$, $\lambda^2=0.1$, disk on both sides]{\includegraphics[scale=0.60]
{KerrWH_lambda01_a070_BS1.eps}\label{fig:finalA}}\hspace{0.3cm}
\subfigure[~Zoomed in version of (a)]{\includegraphics[scale=0.60]{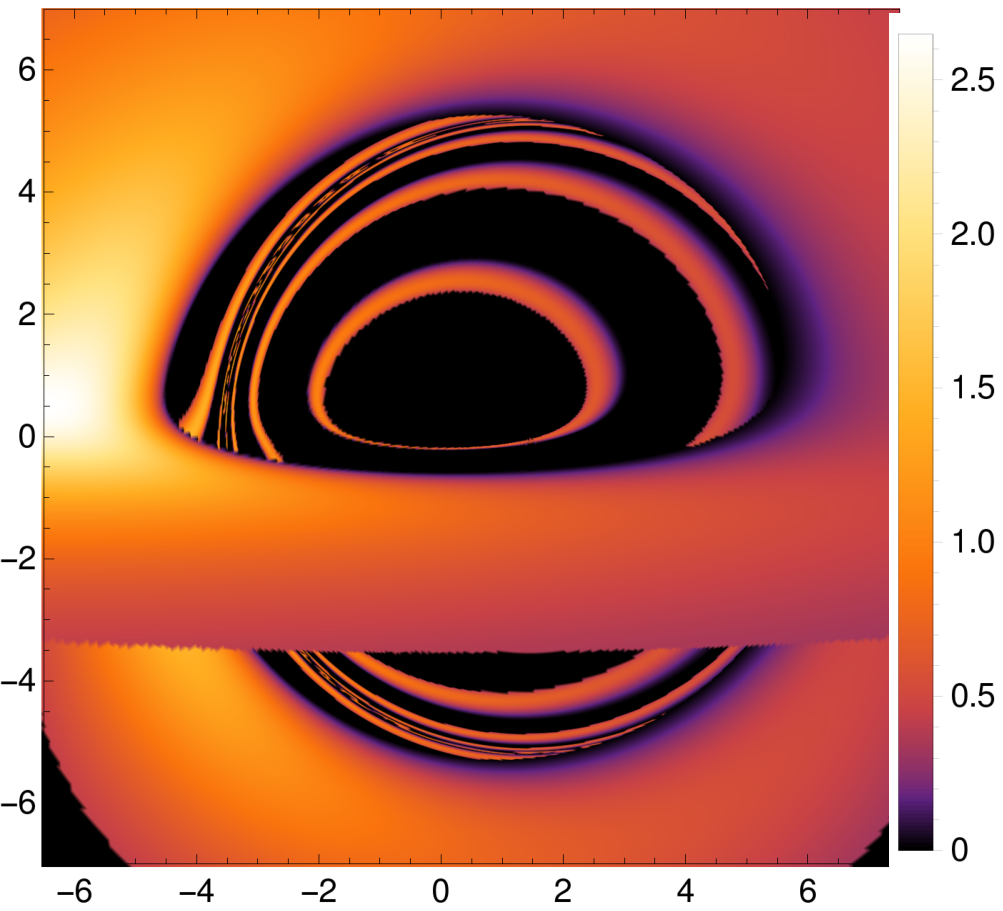}\label{fig:finalB}}
\caption{The images of a Kerr-like wormhole with accretion disks on both sides of the throat [(a)]. (b) is the zoomed in version of (a). The outer edge of the disk is at $r=20M$. The observer's inclination angle is $\theta_{o}=80^{\circ}$. The observer is placed at the radial coordinate $r=10^4M$, which corresponds effectively to the asymptotic infinity. All spatial coordinates are in units of $M$.}
\label{fig:KerrWH_BS}
\end{figure}

Here, we have considered two different classes of rotating wormhole geometries, namely a Kerr-like wormhole
and the Teo wormhole (which is a rotating version of the Morris-Thorne wormhole). We have shown 
that there might be very distinctive features of the accretion disk images from a 
wormhole that can conclusively prove its difference with that in the background of a black hole. 
Specifically, we have shown that this difference arises as the wormhole might have accretion disks 
on both sides of its throat, so that if the observer and the disk are on opposite sides of the
throat, novel images might be obtained which are not known to occur in any other geometry.  
Thus, any accretion disk image that contains such features might be used to distinguish wormholes from other objects. 
It is to be noted that the novel feature (multiple images) of an accretion disk when it is on the other side of the throat are 
generic to wormholes. We have considered two different examples here, but we expect that our analysis is robust, and 
that the features displayed in Fig. (\ref{fig:KerrWH_BS}) will be qualitatively similar in other wormhole geometries.

\end{document}